\documentclass[11pt]{article}
\usepackage[toc,page]{appendix}
\usepackage[utf8x]{inputenc}
\usepackage{setspace}
\usepackage[colorlinks]{hyperref}
\usepackage{amsmath}
\usepackage{amssymb}
\usepackage{amsthm}
\usepackage[frak=esstix]{mathalpha}
\usepackage{mathtools}
\usepackage{physics}
\usepackage{booktabs}
\usepackage{color,soul}
\usepackage{tocloft}
\usepackage{xcolor}
\usepackage{geometry} 
\usepackage{fullpage}[cm] 
\usepackage[font=small,format=plain,labelfont=bf,up,textfont=normal,up]{caption}

\swapnumbers

\selectcolormodel{cmy}
\hypersetup{
	colorlinks,
	linkcolor = {cyan},
	citecolor = {green},
	urlcolor  = {magenta}
}

\newcommand{\diff}{\ensuremath{\textrm{diff}}}
\newcommand{\TT}{\ensuremath{{\textrm{T}\bar{\textrm{T}}}}}
\newcommand{\STT}{\ensuremath{S_{\textrm{T}\bar{\textrm{T}}}}}

\newcommand{\Pd}{\ensuremath{P^{\dagger}}}
\newcommand{\pd}{\ensuremath{\partial}}
\newcommand{\bpd}{\ensuremath{\bar{\partial}}}

\newcommand{\pr}[1]{\left(#1\right)}

\newcommand{\DD}{\ensuremath{\mathcal{D}}}
\newcommand{\OO}{\ensuremath{\mathcal{O}}}

\newcommand{\ATS}{\ensuremath{\mathcal{A}}}
\newcommand{\ee}{\ensuremath{\varepsilon}}

\newcommand{\al}{\ensuremath{\alpha}}
\newcommand{\bb}{\ensuremath{\beta}}
\newcommand{\g}{\ensuremath{\gamma}}
\newcommand{\de}{\ensuremath{\delta}}
\newcommand{\om}{\ensuremath{\omega}}
\newcommand{\Om}{\ensuremath{\Omega}}
\newcommand{\sg}{\ensuremath{\sigma}}
\newcommand{\ep}{\ensuremath{\epsilon}}
\newcommand{\La}{\ensuremath{\Lambda}}
\newcommand{\Th}{\ensuremath{\Theta}}
\newcommand{\ttt}{\ensuremath{\tau}}
\newcommand{\tn}{\ensuremath{\tau_1}}
\newcommand{\tw}{\ensuremath{\tau_2}}

\begin{document}

\vspace{1.8cm}
\begin{center}
  \Huge Correlation Functions in $\TT$-deformed Theories on the Torus
\end{center}  

\vspace{0.7cm}
\begin{center}        
{\large Netanel Barel}
\end{center}

\vspace{0.15cm}
\begin{center}        
\emph{Department of Particle Physics and Astrophysics, Weizmann Institute of Science, \\ Rehovot 7610001, Israel}\\[.4cm]
             
e-mail: \tt netanel.barel@weizmann.ac.il
\end{center}

\vspace{1.5cm}


\begin{abstract}

We study the correlation functions of local operators in unitary $\TT$-deformed field theories defined on a torus, using their formulation in terms of Jackiw-Teitelboim gravity. We focus on the two-point correlation functions in momentum space when the undeformed theory is a conformal field theory. The large momentum behavior of the correlation functions is computed and compared to that of $\TT$-deformed field theories defined on a plane. For the latter, the behavior found was $\pr{\frac{\sqrt{t}|q|}{\pi e}}^{-\frac{tq^2}{\pi}}$, where $q$ is the momentum and $t$ is the deformation parameter. For a torus, the same behavior is found for $|q|<<L/t$, where $L$ is the torus' length scale. However, for $|q|>>L/t$, a different behavior is found: $\pr{\frac{2\sqrt{t}^5q^2}{\pi e L^3|T|^2}}^{\frac{tq^2}{\pi}}$, where $T$ is the complex structure of the torus. Hence, at large momentum, the correlator decays and then grows. This behavior suggests that operators carrying momentum $q$ are smeared on a distance scale $t|q|$. The difference from the plane's result illustrates the non-locality of the theory and the UV-IR mixing. 
 
\end{abstract}
\thispagestyle{empty}


\clearpage

\setstretch{1.35}  
\tableofcontents
\thispagestyle{empty}

\setstretch{1.5}  


\newpage
\setcounter{page}{1}
\section{Introduction}\label{sec_int}
In recent years it was discovered that in 1+1 dimensional space-time, special types of irrelevant deformations of quantum field theories (QFTs) are under control and even solvable \cite{Smirnov:2016lqw,Cavaglia:2016oda}. One of them is the $\TT$-deformation, which is the deformation investigated in this paper. A review of this topic with a complete list of references can be found in \cite{Jiang:2021kfo}. The $\TT$-operator is defined as the determinant of the energy-momentum tensor using the following normalization:
\begin{equation}\label{def_EMT}
    T^{\al\bb} \equiv \frac{2}{\sqrt{g}}\frac{\de S}{\de g_{\al\bb}}\Big|_{g = \de},
\end{equation}
where Euclidean signature is being used throughout this work. $\sg^{1,2}$ are coordinates in the two dimensional space, and $\de$ is the flat metric. The $\TT$-operator is $\TT \equiv \det\pr{ T_{\al\bb} }$.\footnote{In the action this operator is integrated over space. One can show that there is no short-distance singularity when bringing
together the different components of $T_{\al\bb}$.} For each value of the deformation parameter $t$, one defines the deformed Lagrangian at $t+dt$ in terms of the deformed Lagrangian at $t$:
\begin{equation}\label{def_TT_2}
    \mathcal{L}^{(t+dt)} = \mathcal{L}^{(t)} + \TT^{(t)}dt,
\end{equation}
where $\TT^{(t)}$ is obtained using the energy-momentum tensor derived from $S^{(t)}$.\footnote{The coefficients in the definitions of the $\TT$-operator and the flow equation are different from paper to paper, and one has to be careful when comparing quantitative results from different papers.} Only positive $t$ will be considered in this work, for which the deformed theory was found to be unitary.\footnote{At least at large enough volume the energies are real.} 

The energy levels of the theory on a circle, and thus the torus partition function and the S-matrix, are explicitly known for any $t$. However, not much is known about correlators and even what are the appropriate operators in the deformed theory. Using this definition of the deformation, which is, in a sense, perturbative, one can use the unperturbed operators as operators in the deformed theory and calculate their correlation functions.\footnote{Correlation functions of a deformed conformal field theory (CFT) were calculated for small deformation parameter by perturbative methods, see \cite{Kraus:2018xrn, Giribet:2017imm, Aharony:2018vux, He:2020qcs, He:2019vzf, He:2019ahx, Ebert:2020tuy, He:2022jyt, He:2020udl, Hirano:2024eab, He:2023kgq}, with specific attention to the correlation functions of the components of the energy-momentum tensor, see \cite{Rosenhaus:2019utc, Li:2020pwa, Hirano:2020ppu, He:2023hoj}.},\footnote{Another approach to defining operators and correlation functions for J$\bar{\textrm{T}}$-deformed theories was suggested in \cite{Guica:2021fkv, Chakraborty:2023wel}. Other possibilities for defining operators for $\TT$-deformed theories were given in \cite{Kruthoff:2020hsi, Tsolakidis:2024wut}. In \cite{Kruthoff:2020hsi}, the geometry of a cylinder is also discussed in addition to that of a plane.} In particular, the two-point correlation function for large deformation parameter was calculated at large momentum in \cite{Cardy:2019qao} at (4.27) (written in our notations, keeping the leading contribution only)\footnote{Similar result appeared in \cite{Hirano:2020nwq}, using the random geometry approach to the $\TT$-deformation.}:
\begin{equation}\label{eq_cor_Cardy}
    \left(\frac{tq^2}{2\pi}\right)^{\frac{tq^2}{2\pi}}.
\end{equation}

A non-perturbative definition for the $\TT$-deformation was suggested in \cite{Dubovsky:2017cnj}. The suggestion claims that $\TT$-deformed theories can be described using flat space Jackiw-Teitelboim (JT) gravity.\footnote{Using this description, the S-matrix, the spectrum on a circle, and the partition function on the torus were reproduced, see \cite{Dubovsky:2017cnj,Dubovsky:2018bmo}. These calculations confirm this suggestion.} The JT-description also gives a natural suggestion for what the operators in the deformed theory should be. In the previous paper \cite{Aharony:2023dod}, correlation functions of these operators were calculated. Like the result in the perturbative description \eqref{eq_cor_Cardy}, the need for a momentum-dependent renormalization showed up,\footnote{Momentum-dependent renormalization implies that one should not consider the momentum-dependent operator as the Fourier transform of local operator, but rather as fundamental operators.} as well as the independence of the IR CFT data at the UV. However, the behavior of the two-point correlator at large momentum turned out to be very different from the perturbative result \eqref{eq_cor_Cardy}. Rather than exponential growth, it presented an exponential decay\footnote{(115) in \cite{Aharony:2023dod}, with $\mu = \frac{1}{\sqrt{t}}$, see subsection \ref{subsec_mod_cor}. Similar results appeared also in \cite{Cui:2023jrb,Chakraborty:2023wel} using different frameworks.}:
\begin{equation}\label{eq_int_pln}
    \pr{\frac{tq^2}{\pi e}}^{-\frac{t q^2}{2\pi}}.
\end{equation} 
Both results emphasize the non-local nature of the $\TT$-deformation, as the high momentum limit of correlators is not a power law $|q|^{2\al}$. Power-low behavior is a characteristic of local field theories, for which their UV-limit is expected to be a CFT.

The calculation in \cite{Aharony:2023dod} was performed for a theory living on an infinite two-dimensional plane (just ``plane" from now on). In this paper, we calculate the two-point correlator for a theory living on a torus. In a local QFT, for which the UV limit is a CFT, the small distance limit of correlators is the same whether the spacetime is a plane, a torus, or any other two-dimensional Riemannian manifold. However, for non-local theories, global aspects of spacetime may affect local properties, a phenomenon known as UV-IR mixing. Therefore, comparing the torus' correlator to the plane's correlator may reveal another non-local aspect of $\TT$-deformed theories.\footnote{In a CFT, small distances correspond to high momentum, and vice versa. In non-local theories, the last statement is no longer true. In particular, in $\TT$-deformed theory, the fundamental operators are defined in momentum space, and so we compare high momentum behavior rather than small distance behavior.} 

The large momentum limit of the tours' correlator is indeed different than \eqref{eq_int_pln}, and it is found to be:
\begin{equation}\label{eq_int_trs}
    \pr{\frac{2\sqrt{t}^5q^2}{\pi e L^3|T|^2}}^{\frac{tq^2}{\pi}},
\end{equation}
where $L$ is the torus' length scale and $T$ is its complex structure. The main text of this paper contains the detailed calculation leading to \eqref{eq_int_trs} and its physical interpretation. Its structure 
is the following. In section \ref{sec_pre}, we review the JT-formalism and present the general quantities and notations needed for a torus geometry. In section \ref{sec_cal}, we calculate the path integral (PI) leading to a simple integral, out of which the large momentum limit can be evaluated. The evaluation of this PI follows similar lines to the evaluation done in \cite{Aharony:2023dod}. We gauge-fix the diffeomorphism invariance by using the orthogonal gauge for the metric. Then, some formal manipulations will be done on the gauge-fixed expression. Regularization and renormalization procedures will be carried out to make the expression well-defined and finite. In section \ref{sec_sad}, we implement the saddle point approximation to the remaining integral and derive the final result \eqref{eq_int_trs}. Finally, in section \ref{sec_sum}, we will summarize the main steps done in the calculation and discuss the result, in particular, its meaning, its relation to the plane's result, and its relation to other non-local field theories.

Appendix \ref{app_mod} is a reminder of modular transformations, with specific attention to what is needed in the paper. Appendix \ref{app_det} is devoted to determinants appearing during the evaluation of the PI. In appendix \ref{app_fun}, we describe briefly the properties of the $\eta(\ttt)$ and $\Theta(w,\ttt)$ functions, as well as those of the propagator $G(w,\ttt)$. In appendix \ref{app_weyl}, the Weyl anomaly appearing in the calculation is calculated. Finally, in appendix \ref{app_com}, the method of saddle point approximation in the complex plane and its usage in the paper are discussed.


\newpage
\section{Preliminaries}\label{sec_pre}
\subsection{JT-formalism}

The action for the $\TT$-deformed theory in the JT-description is given by:
\begin{equation}\label{eq_per_action1}
\begin{split}
    \STT(\psi,g_{\al\bb},\varphi) &= S_0(\psi,g_{\al\bb}) + S_{JT}(g_{\al\bb},\varphi),
    \\
    S_{JT}(g_{\al\bb},\varphi) &= \int d^2\sg\sqrt{g}\left(\varphi R -\Lambda\right).
\end{split}    
\end{equation}
In this equation, $\psi$ represents the field content of the undeformed theory and $S_0(\psi,g_{\al\bb})$ its action. The action is coupled to a dynamical metric $g_{\al\bb}$, and not to the dilaton field $\varphi$. For a general QFT, the coupling to the metric may be chosen by the minimal coupling. However, in the case where the undeformed theory is a CFT, as in this paper, a more natural coupling is chosen, such that $S_0(\psi,g_{\al\bb})$ admits Weyl invariance.\footnote{The path-integral, however, has a Weyl anomaly, see subsection \ref{subsec_Undeformed-PI} and appendix \ref{app_weyl}.} The energy density in the vacuum $\La$ is connected to the $\TT$-deformation parameter $t$ via:
\begin{equation}\label{INT_5}
    t = -\frac{1}{\Lambda}.
\end{equation}
In this formalism, the unitary deformed theory is found for $\La=-|\La|<0$, and the undeformed theory is recovered in the limit $|\La|\to\infty$.

The JT-gravity action \eqref{eq_per_action1} has an equivalent formulation using vielbein variables $e^a_{\al}$ and a spin-connection $\omega_{\al}$, instead of the metric components $g_{\al\bb} = e^a_\al e^b_\bb \de_{ab}$. One introduces Lagrange multipliers $\lambda^a$ to force the spin-connection $\omega_{\al}$ to be compatible with the metric. Defining $X^a\equiv\Lambda^{-1}\ep^a_b\lambda^b$, simple manipulations on \eqref{eq_per_action1} gives:
\begin{equation}\label{eq_per_action2}
\begin{split}
    \STT(\psi,e^a_{\al},X^a) & = S_0(\psi,e^a_{\al}) + S_{JT}(e^a_{\al},X^a),
    \\
    S_{JT}(e^a_{\al},X^a) & =
    \frac{\Lambda}{2}\int d^2\sg \ep^{\al\bb}\ep_{ab}(\partial_{\al}X^a-e^a_{\al})(\partial_{\bb}X^b-e^b_{\bb}).
\end{split}
\end{equation} 
The fields $X^a(\sg)$ play the role of dynamical coordinates of the target-space (TS), which is identified with the space of the undeformed theory \cite{Dubovsky:2017cnj,Dubovsky:2018bmo}, i.e. a torus in the present paper. The world-sheet (WS), on which the actions (\ref{eq_per_action1},\ref{eq_per_action2}) are formulated, has the same topology as the TS, so it is also a torus. The dynamical coordinates describe one winding of the WS torus (WSt) around the TS torus (TSt). Using these dynamical coordinates, local operators can be defined in the deformed theory:
\begin{equation}\label{op_def_pos}
    \OO(X) \equiv \int d^2\sg\sqrt{g(\sg)} O(\sg)\de\left(X(\sg)-X\right),
\end{equation} 
where $O(\sg)$ is some operator in the undeformed theory.\footnote{We consider scalar operators for simplicity. If the undeformed operator has a spin, additional vielbein should be inserted. For example, for the operator $O^\al(\sg)$, one defines: $\OO^a(X) = \int d^2\sg\sqrt{g(\sg)} e^a_\al(\sg)O^\al(\sg)\de\left(X(\sg)-X\right)$. The vielbein changes the WS index to a TS index and keeps the diffeomorphism invariance of the deformed operator.} As it turns out, the Fourier transform of these operators is more natural to use:
\begin{equation}\label{op_def_mom}
    \OO(q) \equiv \int d^2X e^{iqX} \OO(X) =  \int d^2\sg\sqrt{g(\sg)} O(\sg)\exp\left(iq\cdot X(\sg)\right).
\end{equation}
In particular, $\OO(q)$ will be multiplied with momentum-dependent renormalization factor (see \eqref{UV_6}), so they should not be viewed as a Fourier transform of the position space operators, but rather as fundamental operators, out of which spatial operators can be defined.

The correlation function in the JT-formalism is defined as follows:
\begin{equation}\label{def_cor}
\begin{split}
    C(q_1,q_2) \equiv& \left<\OO(q_1)\OO(q_2)\right> = \frac{1}{Z_{\TT}}\int\frac{\DD e\DD X\DD\psi}{V_\diff}\OO(q_1)\OO(q_2)e^{-\STT}, 
    \\
    Z_{\TT} \equiv & \int\frac{\DD e\DD X\DD\psi}{V_\diff}e^{-\STT}.
\end{split}
\end{equation}
The partition function $Z_{\TT}$ was already calculated in \cite{Dubovsky:2018bmo}. It will be omitted from now on since it does not depend on the momentum. Plugging the definition of the deformed operators \eqref{op_def_mom} inside \eqref{def_cor} gives:
\begin{equation}\label{eq_cor_dev_1}
    C(q_1,q_2) = \frac{1}{Z_{\TT}}\int\frac{\DD e\DD X\DD\psi}{V_\diff}\int d^2\sg_1d^2\sg_2 \sqrt{g(\sg_1)}\sqrt{g(\sg_2)}O(\sg_1)O(\sg_2)e^{iq_1\cdot X(\sg_1)+iq_2\cdot X(\sg_2)-\STT}.
\end{equation}


\subsection{Torus' Notations}\label{subsec_torus_notation}

For the WSt we present periodic Cartesian coordinates:
\begin{equation}\label{eq_sg_domain}
    \sg^1 \sim \sg^1 + 1, \ \sg^2 \sim \sg^2 + 1.
\end{equation}
The line element in the orthogonal gauge depends on the conformal factor $e^{\Om}$ and the complex structure of the WSt $\ttt \equiv \tn + i\tw$:
\begin{equation}
    ds^2 = e^{2\Om}\pr{(d\sg^1 + \tn d\sg^2)^2 + \pr{\tw d\sg^2}^2} = e^{2\Om}|d\sg^1 + \ttt d\sg^2|^2.
\end{equation}   
From the line element, one reads the metric:
\begin{equation}
\begin{split}
    [\bar g_{\al\bb}]_\sg & =
    e^{2\Om}
    \begin{pmatrix} 
    1 & \tn \\ \tn & |\ttt|^2 \end{pmatrix}
    = e^{2\Om}[\bar {g_0}_{\al\bb}]_\sg,
    \\
    [\pr{g_0}_{\al\bb}]_\sg & = M^TM, 
    \\
    M &\equiv \begin{pmatrix} 1 & \tn \\ 0 & \tw \end{pmatrix}.
\end{split}
\end{equation}
$[\pr{g_0}_{\al\bb}]_\sg$ is the flat metric of the undeformed theory. The vielbein is given by:
\begin{equation}\label{eq_vilebein_sg}
\begin{split}
    [\bar e^a_\al]_\sg =      
    e^{\Om + \ep\phi}M
    =
    e^{\Om}
    \begin{pmatrix} 
    \cos(\phi) & \tn\cos(\phi) + \tw\sin(\phi) \\ -\sin(\phi) & -\tn\sin(\phi) + \tw\cos(\phi)
    \end{pmatrix},
\end{split}
\end{equation}    
where $\ep_{ab}$ is the anti-symmetric symbol, $\ep_{12} = - \ep_{21} = 1, \ep_{11} = \ep_{22} = 0$. The WS area element is $d\bar A = \sqrt{g} d^2\sg = e^{2\Om}\tw d^2\sg$. After gauge-fixing the diffeomorphism invariance, all possible $\Om(\sg),\phi(\sg)$ and $\ttt$ will be integrated.

It is convenient to present the complex coordinates:
\begin{equation}
    w \equiv \sg^1 + \ttt\sg^2, \ \bar w \equiv \sg^1 + \bar\ttt\sg^2.
\end{equation}
The line element in these coordinates is $ds^2 = dwd\bar w$, and the metric and the vielbein are given by:
\begin{equation}
\begin{split}
    [\bar g_{\al\bb}]_w & =      
    e^{2\Om}
    \begin{pmatrix} 
    0 & \frac{1}{2} \\ \frac{1}{2} & 0 \end{pmatrix},
    \\
    [\bar e^a_\al]_w & =      
    e^{\Om + \ep\phi}
    \begin{pmatrix} 
    \frac{1}{2} & \frac{1}{2} \\ -\frac{i}{2} & \frac{i}{2} 
    \end{pmatrix}. 
\end{split}
\end{equation}
Some useful relations to the Cartesian coordinates that will be used all along the paper and are given here:
\begin{equation}
\begin{split}
    d^2w &= 2\tw d^2\sg,
    \\
    \de(w) &= \frac{1}{2\tw}\de(\sg),
    \\
    \begin{pmatrix}
        \pd \\ \bpd 
    \end{pmatrix}
    &= \frac{i}{2\tw}
    \begin{pmatrix}
        \bar\ttt & -1 \\ -\ttt & 1 
    \end{pmatrix}
    \begin{pmatrix}
        \pd_1 \\ \pd_2 
    \end{pmatrix},
    \\
    4\pd\bpd & = \frac{1}{\tw^2}\pr{|\ttt|^2\pd_1^2 + \pd_2^2 -2\tn\pd_1\pd_2} = e^{2\Om} g^{\al\bb}\pd_\al\pd_\bb.
\end{split}
\end{equation}

We turn to the TSt. Two vectors $L_\al$ specifying it can be written as:
\begin{equation}\label{eq_vielbein_TSt}
    {L^a}_\al = 
    \begin{pmatrix} 
    X & XT_1 + YT_2 \\ 
    -Y & -YT_1 + XT_2
    \end{pmatrix}.
\end{equation}
They are constant vielbein in the TS. $X,Y$ contains the information about the length and orientation of $L_1$.\footnote{In the WSt, instead of the ``Cartesian" variables $X,Y$, polar variables $e^\Om,\phi$ were used for the object $e^a_\al$. On the WSt $e^a_\al$ change locally, while on the flat TSt they can be defined globally} $T \equiv T_1 + iT_2$ is the modular parameter of the TSt. 

The momenta $q$ appearing in the Fourier transform on the TSt, maintain the periodicity $X^a \sim X^a + L^a_\al$:
\begin{equation}
    e^{q_aL^a_\al} = 1 \Rightarrow q_aL^a_\al = 2\pi n_\al,
\end{equation}
for some integers $n_\al$. Momenta satisfying this condition are given by the reciprocal lattice vectors: 
\begin{equation}
\begin{split}
    {\pr{L^{-1}}^\al}_a & = \frac{1}{\ATS}
    \begin{pmatrix} 
    -YT_1 + XT_2 & -XT_1 - YT_2 
    \\ Y & X
    \end{pmatrix},
    \\
    q_a & = 2\pi n_\al \pr{L^{-1}}^\al_a.
\end{split}    
\end{equation}
One can choose $L_1$ to be along the x-axis $X = L,Y=0$, leading to:
\begin{equation}
    q = \frac{2\pi}{L}\pr{n_1\pr{1,-\frac{T_1}{T_2}} + n_2\pr{0,\frac{1}{T_2}}} = \frac{2\pi}{L}\pr{n_1,\frac{-n_1T_1 + n_2}{T_2}}.    
\end{equation}

The transformation $X^a(\sg) = L^a_\al\sg^\al$ maps the WSt to the TSt, wrapping the last once. Many such different mappings are generated by this map by applying a modular transformation on $T$ or $\ttt$, but not on both. Applying it for one of them, although not changing the torus, changes the two TS cycles into which the two WS cycles $\sg^1,\sg^2$ are mapped. Applying the same modular transformation to $\ttt$ and $T$ gives an identical map, and it is a symmetry of the action \eqref{eq_per_action2}. We will fix one modular parameter $T$, and allow all possible modular parameters $\ttt$ in the upper half plane. We refer the reader to appendix \ref{app_mod} for more details.


\newpage
\section{PI-Calculation}\label{sec_cal}
The major part of this section is technical, and it is similar to section 3 of \cite{Aharony:2023dod}. The main difference is in subsection \ref{subsec_vielbein} regarding the vielbein solution and, consequently, the expression for the WS-area. The reader who is not interested in the technical details can read only this subsection, and skip to the next section.

\subsection{Gauge-Fixing}\label{subsec_gauge}

Our starting point is \eqref{eq_cor_dev_1}:
\begin{equation}\label{eq_main_1}
    \int\frac{\DD e\DD X\DD\psi}{V_\diff}\int d\sg_1d^2\sg_2 \sqrt{g(\sg_1)}\sqrt{g(\sg_2)}O(\sg_1)O(\sg_2)e^{iq_1\cdot X(\sg_1)+iq_2\cdot X(\sg_2)-\STT}.
\end{equation}
The first stage is to gauge-fix the diffeomorphism invariance. We choose the orthogonal gauge for the vielbein, written in subsection \ref{subsec_torus_notation}. There are two remaining transformations not fixed by this choice. These are translations of $\sg^\al$, which are isometries of the gauged-fixed metric. We fix them further by placing one of the operators at $\sg_1^\al = 0$. The FP-identity takes the form:
\begin{equation}\label{eq_FP_identity}
    1 = \int\DD V(\sg)\DD\Om(\sg)\DD\phi(\sg) d^2\ttt J(e,\sg_1) \de\pr{ e^{(V)} - \bar e(\Om(\sg),\phi(\sg),\ttt) }\de\left(\sg_1^{(V)}-0\right).
\end{equation}
The determinant $J$ is hard to calculate; see appendix \ref{app_det}. However, it will not be relevant for the high momentum limit, as we claim in that appendix. It is also reasonable from another point of view, as the determinant is a sense a local property, and expected to be similar to the determinant calculated for the plane. The latter, indeed, was independent of the momentum at high momentum.

Inserting the FP-identity into \eqref{eq_main_1} and performing the usual FP-manipulations gives:
\begin{equation}\label{eq_main_2}
    \int \DD X \DD\psi\DD\Om\DD\phi d^2\ttt d\bar A(\sg_1)d\bar A(\sg_2)\de(\sg_1) J(\bar{e},0) O(\sg_1)O(\sg_2)e^{iq_1\cdot X(\sg_1)+iq_2\cdot X(\sg_2) -S_0 -S_{JT}}.
\end{equation}


\subsection{Target-Space Coordinate PI}

To evaluate the PI over $X^a(\sg)$, we decompose it into a ``one winding" mode (see subsection \ref{subsec_torus_notation}) and all other modes orthogonal to it:
\begin{equation}\label{rel_X_Y}
    X^a(\sg) = L^a_\al\sg^\al + Y^a(\sg).
\end{equation}
$Y^a(\sg)$ are periodic functions on the WSt. Plugging this decomposition into the $X$-dependent terms in \eqref{eq_main_2} reads:
\begin{equation}
    e^{2\pi in_{1\al}\sg_1^\al + 2\pi in_{2\al}\sg_2^\al } \int\DD Y e^{iq_1\cdot Y(\sg_1) + iq_2\cdot Y(\sg_2) + \La\int d^2\sg \ep^{\al\bb}\ep_{ab}\bar{e}^a_{\al}(\sg)\pd_{\bb}Y^b(\sg)},
\end{equation}
where $n_{i\al} \equiv \pr{q_i}_aL^a_\al$ are integers. The remaining terms in $e^{-\STT}$ which are independent of $Y$ are:
\begin{equation}
    e^{|\La|\bar A} e^{|\La|\ATS}
    e^{\La\int d^2\sg \ep^{\al\bb}\ep_{ab}L^a_{\al}\bar{e}^b_{\bb}}.
\end{equation}
The $\DD Y$ integration is ill-defined because the exponent is not a pure phase. It will become a pure phase by Wick rotating $\bar{e}$, i.e. $e^{\Om}\rightarrow ie^{\Om}$. In terms of $\Om,\phi$, it sends $\Om\rightarrow\Om+i\pi/2$, and does not change the measure $\DD\Om\DD\phi$. The $\DD Y$-integration becomes trivial and gives a $\de$-function. We use a discretized notation to take care of the $\de$-functions appearing, which include momentum conservation. The discretized measure is $\DD Y = \det\pr{e^{2\Om(\sg)}\tw}\displaystyle \prod_{b=1}^2\prod_{\sg=1}^N dY^b(\sg)$, and the $\DD Y$-integration yields:
\begin{equation}\label{eq_main_3}
\begin{split}
    e^{|\La|\ATS} \de(q_1+q_2)\int &\DD\psi\DD\Om\DD\phi d^2\ttt d\bar A(\sg_1)d\bar A(\sg_2)\de(\sg_1) J(\bar{e},0) \det\left(e^{2\Om}\tw\right)\prod_{b=1}^2 \prod_{\sg=1}^{N-1}\de(K_b(\sg)) \cdot \\
    &  O(\sg_1)O(\sg_2) e^{2\pi in_{1\al}\sg^\al_1 + 2\pi in_{2\al}\sg^\al_2 - S_0(\psi,\bar{e}) - |\La|\pr{\bar{A} + i\int d^2\sg \ep^{\al\bb}\ep_{ab}L^a_{\al}\bar{e}^b_{\bb}}},
    \\
    K_b(\sg) \equiv & q_{1b}\de_{\sg,\sg_1} + q_{2b}\de_{\sg,\sg_2} - \La \ep^{\al\bb}\ep_{ab}\pd_{\bb}\bar{e}^a_{\al}(\sg).
\end{split}
\end{equation}

We show it for the circle of length $L$ for simplicity. The analog of our integral is:
\begin{equation}\label{eq_Dy_circle}
    \int_0^L dY(1) \int_{-\infty}^\infty \prod_{i=2}^N dY(i) e^{i\sum_{j=1}^N f_jY(j)}.
\end{equation}
The periodicity of $Y$ is implemented by the integration domain of $Y(1)$, from $0$ to $L$. All the rest are not limited anymore. $f_j$ contains the derivative of the vielbein plus the momenta $\frac{2\pi n_j}{L}$ ($n_j=0$ for all $N-2$ locations without momentum insertions). By changing variables $Y(i) = Y(1) + Z(i)$ for $2\leq i \leq N$, the integral \eqref{eq_Dy_circle} becomes:
\begin{equation}
    \int_0^L dY(1) e^{iY(1)\sum_{j=1}^N f_j} \int_{-\infty}^\infty\prod_{i=2}^N dZ(i) e^{i\sum_{j=2}^N f_jZ(j)}.
\end{equation}
Since the derivative of the vielbein vanishes upon summation, $\sum_{j=1}^N f_j = \frac{2\pi}{L}\sum_{j=1}^N n_j$. The integral over $Y(1)$, although not extended to infinity, gives a Kronecker-delta:
\begin{equation}
    L\de_{0,\sum_{j=1}^N n_j}\prod_{i=1}^{N-1}2\pi\de(f_j) = \det(2\pi)\de\pr{\sum_i q_i} \prod_{j=1}^{N-1}\de(f_j),
\end{equation}
which is the analog of \eqref{eq_main_3}. In the last line we used the $\de$-functions, $\sum_{i=1}^N f_i = 0$ and $f_i = 0$ for $2\leq i \leq N$, and a continuous $\de\pr{\sum_i q_i}$ instead of the Kronecker-delta. The infinite factor $\det(2\pi)^2$ was omitted.


\subsection{Conformal-factor and Orientation PI}\label{subsec_conf-ori-PI}

The integral $\DD\Om\DD\phi$ is easily performed because of the constraint $K_b(\sg) = 0$. The constraint almost completely restricts the possible $\Om(\sg),\phi(\sg)$. Given one solution $\Om_0,\phi_0$, all other solutions $f_\Om,f_\phi$ are given by:
\begin{equation}
\begin{split}
     e^{f_{\Om} + \ep f_{\phi}} & \equiv e^{\Om_0 + \ep\phi_0} + e^{\bar\Om + \ep\bar\phi} \Rightarrow 
     \\
    f_{\Om}(\Om_0,\bar{\Om},\phi_0,\bar{\phi}) & \equiv \frac{1}{2}\ln\left(e^{2\Om_0}+e^{2\bar{\Om}}+2e^{\Om_0+\bar{\Om}}\cos(\phi_0-\bar{\phi})\right),
    \\
    f_{\phi}(\Om_0,\bar{\Om},\phi_0,\bar{\phi}) & \equiv \tan^{-1}\left(\frac{e^{\Om_0}\sin(\phi_0)+e^{\bar{\Om}}\sin(\bar{\phi})}{e^{\Om_0}\cos(\phi_0)+e^{\bar{\Om}}\cos(\bar{\phi})}\right),
\end{split}
\end{equation}
where $\bar\Om,\bar\phi$ are constants. We discretize the $\DD\Om,\DD\phi$ measure as $\det\pr{e^{2\Om(\sg)}\tw} \displaystyle \prod_{\sg=1}^N d\Om(\sg) d\phi(\sg)$. Out of the infinite integrals over $\Om(\sg),\phi(\sg)$, only two over $\bar\Om,\bar\phi$ remains. Separating them properly from  the measure, for a general functional $\mathcal{F}$, gives:
\begin{equation}\label{OP_PI_5}
    \int \pr{\prod_{\sg=1}^N d\Om(\sg) d\phi(\sg)} \prod_{b=1}^2\prod_{\sg'=1}^{N-1} \de(K_b(\sg'))\mathcal{F}(\Om(\sg),\phi(\sg)) = \int d\bar{\Om}d\bar{\phi} \sqrt{G^{\bar{\Om},\bar{\phi}}} \frac{\mathcal{F}(f_{\Om},f_{\phi})}{\sqrt{{\det}'(Q^{\dagger}Q)}}.
\end{equation} 
In \eqref{OP_PI_5}, $Q(f_{\Om},f_{\phi})$ denotes the linearization of the constraints $K_a(\sg)$ around the solution $f_{\Om},f_{\phi}$, $\det'$ excludes the two zero modes left to be integrated (the kernel of $Q$), and $\sqrt{G^{\bar{\Om},\bar{\phi}}}$ is the volume element in this two dimensional zero subspace. $\sqrt{G^{\bar{\Om},\bar{\phi}}}$ and $Q^{\dagger}$ are calculated in appendix \ref{app_det}. They are given by:
\begin{align}
    \sqrt{G^{\bar{\Om},\bar{\phi}}} & = \int d^2\sg e^{2\bar\Om-2f_\Om},
    \\
    \sqrt{{\det}'(Q^{\dagger}Q)} & = \det(e^{2\Om}\tw) \frac{\int d^2\sg e^{-2f_\Om}}{\tw} \left|{\det}'\pr{\square}\right|,
    \\
    \square & \equiv -\frac{1}{\tw}(\pd +\pd\ln(\bar z))(\bpd + \bpd\ln(\bar z)).
\end{align}
Using \eqref{OP_PI_5} inside \eqref{eq_main_3} yields:
\begin{equation}\label{eq_main_4}
\begin{split}
    e^{|\La|\ATS}\de(q_1+q_2) \int & \DD\psi d^2\ttt d\bar{\Om}d\bar{\phi} d\bar A(\sg_1)d\bar A(\sg_2)\de(\sg_1) \frac{\det(e^{2f_\Om}\tw) J(\bar{e},0) e^{2\bar\Om}\tw}{\left|{\det}'\pr{\square}\right|} \cdot \\
    & O(\sg_1)O(\sg_2) e^{2\pi in_{1\al}\sg^\al_1 + 2\pi in_{2\al}\sg^\al_2 - S_0(\psi,\bar{e}) - |\La|\pr{\bar{A} + i\int d^2\sg \ep^{\al\bb}\ep_{ab}L^a_{\al}\bar{e}^b_{\bb}}}.
\end{split}
\end{equation}
Since we performed the $\DD\Om,\DD\phi$ integration, $f_\Om,f_\phi$ will be denoted simply by $\Om,\phi$.


\subsection{Vielbein solution}\label{subsec_vielbein}

$K_b(\sg) = 0$ are two differential equations for the vielbein:
\begin{equation}
\begin{split}
    \pd_1{e^2}_2 - \pd_2{e^2}_1 & = - \sum_{i=1}^2 \frac{(q_i)_1}{|\La|}\de(\sg - \sg_i),
    \\
    -\pd_1{e^1}_2 + \pd_2{e^1}_1 & = - \sum_{i=1}^2 \frac{(q_i)_2}{|\La|}\de(\sg - \sg_i).
\end{split}
\end{equation}
Adding the second equation multiplied by $i$ to the first equation gives:
\begin{equation}
\begin{split}
    (-i\ttt\pd_1 + i\pd_2)\bar z & = -2\pi \sum_{i=1}^2 Q_i\de(\sg - \sg_i) \Rightarrow
    \\
    \bpd\bar z & = -2\pi \sum_i Q_i\de(w - w_i).
\end{split}
\end{equation}
where:
\begin{equation}
\begin{split}
    \pr{Q_i}_a & \equiv \frac{(q_i)_a}{2\pi|\La|}, \\
    Q_i & \equiv \pr{Q_i}_1 + i\pr{Q_i}_2, \\
    z & \equiv e^{\Om + i\phi}.
\end{split}
\end{equation}
We solve this equation using the Green function for the torus:
\begin{align}\label{eq_def_G}
    \pd\bpd G & = -2\pi \de(w) + \frac{\pi}{\tw}, 
    \\
    G & = -\ln\left| \Th\pr{w,\ttt} \right|^2 + \frac{2\pi\Im(w)^2}{\tw},
\end{align}
where $\Th\pr{w,\ttt}$ is one of the Jacobi Theta functions.\footnote{The first equation can be written in a diffeomorphism invariant way: $g^{\al\bb}\pd_\al\pd_\bb G = -4\pi\de_g + \frac{4\pi}{\bar A}$, with $\de_g \equiv \frac{\de^2(\sg)}{\sqrt{g}}$. The second term is a uniform charge density, making the total charge vanish. This is a consequence of the Gauss law used in compact spaces.} Solving for each source $\de(w-w_i)$, and multiplying by $Q_i$, gives $\bar z$:
\begin{equation}\label{eq_z_sol}
\begin{split}
    Q_i\pd\bpd G_i & = -2\pi Q_i\de(w - w_i) + \frac{\pi Q_i}{\tw} \Rightarrow 
    \\
    \bar z & = \sum_i Q_i \pd G_i + e^{\bar\Om + i\bar\phi}.
\end{split}
\end{equation}
where in the last equation we used momentum conservation $\sum_i Q_i = 0$.\footnote{Since $\sg,\ttt,w$ are dimensionless, $e^{\bar\Om + \bar\phi}$ should have mass dimension $-1$, like $Q_i$. One could either multiply it by $L$ or declare $R=e^{\bar\Om}$ to be dimensionful. The notation $e^{\bar\Om}$ is not appropriate then, but we will use it for convenience.} In the terminology of subsection \ref{subsec_conf-ori-PI}:
\begin{equation}
    e^{\Om_0 + i\phi_0} \equiv \sum_i Q_i \pd G_i.
\end{equation}
The solution $G$ is not unique; any function $f(\ttt,\bar\ttt)$ can be added to it. Also, $G$ is not modular invariant - it is not invariant under $\ttt \to -1/\ttt$; rather, it shifts by $-\ln|\ttt|$. One can modify it by using $\eta$-function: 
\begin{equation}
    G' \equiv G + \ln|\eta(\ttt)|^2 = -\ln\left| \frac{\Theta\pr{w,\ttt}}{\eta(\ttt)}\right|^2 + \frac{2\pi\Im(w)^2}{\tw}.
\end{equation}
$G'$ is modular invariant:
\begin{equation}
    G'(w,\ttt+1) = G'(w,\ttt), \ G'\pr{\frac{w}{\ttt},-\frac{1}{\ttt}} = G'(w,\ttt).
\end{equation}
However, momentum conservation implies $\sum_{i=1} Q_iG'_i = \sum_{i=1} Q_iG_i$, and the same for all of the derivatives of $G'$, so we will use $G$ instead. Basic definitions and properties of $\Th,\eta$-functions and $G$ are given in appendix \ref{app_fun}.

The WS-area is:
\begin{equation}
    \bar A = \int d^2\sg \sqrt{\bar g} = \int d^2\sg e^{2\Om}\tw = \int \frac{d^2w}{2} |z|^2.
\end{equation}
Using \eqref{eq_z_sol} one can write $|z|^2$:
\begin{equation}
    |z|^2 = \sum_{ij} Q_i\bar Q_j \pd G_i \bpd\bar G_j + \sum_i 2\Re(Q_i\pd G_i e^{\bar\Om - i\bar\phi}) + e^{2\bar\Om}.
\end{equation}
The integral over $|z|^2$ can be evaluated easily using integration by parts, the EOM of $G$, and momentum conservation:
\begin{equation}
\begin{split}
    \int d^2w \pd G_i = \int d^2w \bpd G_i = 0 & \ \text{(periodicity)},
    \\
    \sum_{ij}Q_i\bar Q_j \int d^2w G_i = \sum_j \bar Q_j (\dots) = 0  & \ \text{(momentum conservation)}.
\end{split}
\end{equation}
Then:
\begin{equation}\label{eq_WSarea}
\begin{split}
    \bar A & = \pi\sum_{ij}Q_i\bar Q_jG_i(w_j) + e^{2\bar\Om}\tw
    \\
    & = \frac{1}{4\pi|\La|^2}\sum_i q_i^2G_{ii} + \frac{1}{2\pi|\La|^2}\sum_{i<j}G_{ij}q_i\cdot q_j + e^{2\bar\Om}\tw.
\end{split}    
\end{equation}
In the last line we used $\bar Q_i Q_j + \bar Q_j Q_i = \frac{2}{(2\pi|\La|)^2}q_i\cdot q_j$ and defined $G_{ij} \equiv G_i(w_j) = G(w_{ij})$.
We note that $\tw$ is absent from the first two terms in the R.H.S. of \eqref{eq_WSarea}, since $\int d^2w\de(w) = 1$, and that the sum reduces to $\frac{q^2}{2\pi|\La|^2}\pr{G_{11} - G_{12}}$. 

The divergent pieces of $\bar A$, $G_{ii}$, arise from the divergence of 
$G$ at $w=0$. $G$ at $|w|<<1$ can be approximated by (see \eqref{eq_Theta_small_w}): 
\begin{equation}\label{eq_G_short}
    G(w,\ttt) \approx -\ln(2\pi)^2 -\ln|\eta(\ttt)|^6 -\ln|w|^2.
\end{equation}
The logarithmic divergence appeared also in the plane. It is expected since the solution of \eqref{eq_def_G} at small $w$ will not depend on the global properties of the WS. This divergence will be treated by point-splitting regularization. However, it is not trivial on which coordinates this splitting should be implied. There are two natural candidates - the Cartesian coordinates and the complex coordinates. It turns out that one should take $|w| = \ep$, rather than $|\sg| = \ep$.\footnote{For the plane, the complex coordinate was $w = \sg^1 + i\sg^2$. The analog of $G$ here, solved the equation $\pd\bpd G = -2\pi\de(w)$, with the solution $-\ln\pr{|w|^2}$. The same sum in \eqref{eq_WSarea} appeared for the plane ((97) in \cite{Aharony:2023dod}), with $G_{ij} \to -\ln|w_{ij}|^2$. There $|w|=|\sg|$, and $-\ln|w_{ii}|^2$ was simply regularized as $-\ln(\ep^2)$.} It has the following advantages:
\begin{itemize}
    \item it makes the final result finite,
    \item the plane's result is recovered in the appropriate limit,
    \item the normalization of the operators is the as for the plane,
    \item and the regularized WS-area is always positive, in particular, when $\tw \to \infty$. We illustrate this below.
\end{itemize}  
One disadvantage is that the regularized WS-area is not modular invariant, i.e. $\ln{|\eta(\ttt)|^6} + G(w_{12},\ttt)$ is not modular invariant (while $\ln{|\eta(\ttt)|^2} + G(w_{12},\ttt)$ is).\footnote{Since $PSL(2,\mathbb{Z})_{WS}$ are part of the diffeomorphism invariance, they can not be broken. The regularization is chosen for one $\ttt$, and for other $\ttt$ obtained by $PSL(2,\mathbb{Z})_{WS}$ transformation, it will be chosen in a way such that the WS-area remains the same. So the exact statement is that the regularization transforms in a complicated way under $PSL(2,\mathbb{Z})_{WS}$. However, the global $PSL(2,\mathbb{Z})_{TS}$ is broken by this regularization; see appendix \ref{app_mod}.}

Therefore, we use $G_{ii} \to -\ln\pr{\ep^2} - \ln\pr{|\eta(\ttt)|^6}$, and the regularized area is:
\begin{equation}
    \bar A = -\frac{\ln\pr{\ep^2}}{4\pi|\La|^2}\sum_i q_i^2 -\frac{\ln\pr{|\eta(\ttt)|^6}}{4\pi|\La|^2}\sum_i q_i^2 + \frac{1}{2\pi|\La|^2}\sum_{i<j}G_{ij}q_i\cdot q_j + e^{2\bar\Om}\tw.
\end{equation}
The finite part for our two-pt correlator reads: \begin{equation}
    -\frac{q^2}{2\pi|\La|^2}\pr{\ln{|\eta(\ttt)|^6} + G_{12}} + e^{2\bar\Om}\tw,
\end{equation}
implementing momentum conservation.

The remaining $\tw$ integral is finite with this regularized area. It would not be so without the additional $- \ln\pr{|\eta(\ttt)|^6}$. At large momentum, the integrand in \eqref{eq_main_4} will be dominated by $e^{-|\La|\bar A}$, and the integral should converge since $\bar A$ should be positive. However, when we regularize and pull out the divergent piece $-\ln\pr{\ep^2}$, the remaining piece might be negative, as for $\tw \gtrsim 2$ (see \eqref{eq_large_tw_G}):
\begin{equation}
    G(w,\ttt) \approx 2\pi\pr{\sg^2-\frac{1}{2}}^2\tw = \frac{\pi}{2}\pr{1- \frac{\sg^2}{\pi}}^2\tw.
\end{equation}
The finite piece $-\ln|\eta(\ttt)|^6$ we added in $G_{ii}$, at $\tw \gtrsim 2$ behaves as $\frac{\pi}{2}\tw$. Then the integrand $e^{-|\La|\bar A}$ is approximately:
\begin{equation}
    e^{-\frac{q^2}{2\pi|\La|}(G_{11}-G_{12})} \underset{\tw \to \infty}{\approx} 
    e^{-\frac{q^2}{2\pi|\La|}\pr{1-{\pr{1- 2\sg^2}^2}}\frac{\pi\tw}{2}},
\end{equation}
and the integral is now convergent.\footnote{However, since an analytic continuation is made for the $\sg$ integral in \eqref{eq_main_4}, it might be not relevant, as will turn out to be true for the large $\tw$ behavior.\label{re_large_tau}}

Another quantity that should be regularized at this stage is the area element $\sqrt{g(\sg_i)}$. We use point-splitting again $|w_i-w_i| \to \ep$, and therefore:
\begin{align}
    e^{\Om(\sg_i)} = |z(w_i)| &\to \frac{|Q_i|}{\ep},
    \\
    \sqrt{g(\sg_i)} = |z(w_i)|^2\tw &\to \frac{|Q_i|^2}{\ep^2}\tw.
\end{align}

The explicit form of the vielbein can be used to simplify the last integral in the exponent of \eqref{eq_main_4}:
\begin{equation}\label{eq_remining_integral}
    -i|\La|\int d^2\sg \ep^{\al\bb}\ep_{ab}L^a_{\al}\bar{e}^b_{\bb}.
\end{equation}
Using \eqref{eq_z_sol} the vielbein can be written as:
\begin{equation}
\begin{split}
    e^a_\al & \equiv \pr{e_s}^a_\al + \pr{e_c}^a_\al 
    \\
    & \equiv e^{\Om_0 + \ep\phi_0}\begin{pmatrix} 1 & \tn \\ 0 & \tw \end{pmatrix} 
    + 
    e^{\bar\Om + \ep\bar\phi}\begin{pmatrix} 1 & \tn \\ 0 & \tw \end{pmatrix}.
\end{split}    
\end{equation}

The coordinate dependent part of the vielbein $\pr{e_s}^a_\al$ does not contribute to \eqref{eq_remining_integral} since $G_i$ is periodic:
\begin{equation}
    \int d^2\sg e^{\Om_0}cos(\phi_0) = \frac{1}{2}\sum_i \int d^2\sg (Q_i\pd G_i + c.c.) = 0.     
\end{equation}
The same it true for $\int d^2\sg e^{\Om_0}sin(\phi_0)$. Therefore only the constant part $\pr{e_c}^a_\al$ contributes, and the integral is trivial: $-i|\La|\ep^{\al\bb}\ep_{ab}L^a_{\al}\pr{e_c}^b_{\bb}$.

\subsection{Undeformed Fields PI}\label{subsec_Undeformed-PI}

Until now we did not use the fact that the undeformed theory is a CFT. From here on, we will use this assumption. For a primary operator $O(\sg)$ with a conformal dimension $\Delta$, the PI over the undeformed fields is:
\begin{equation}\label{P_PI_3}
     \int\DD\psi O(\sg_1)O(\sg_2) e^{-S_0\left(\psi,e^{2\Om}g_0\right)} =
     e^{-(\Om(\sg_1)+\Om(\sg_2))\Delta +W}Z_0[g_0]C^0(|\sg_1-\sg_2|),
\end{equation}
where $Z_0[g_0],C^0(|\sg_1-\sg_2|)$ are the undeformed partition function and correlator respectively. $e^W$ is the Weyl anomaly:
\begin{equation}\label{eq_weyl_1}
    W = \frac{c}{12}\int d^2\sg \sqrt{g_0}g_0^{\al\bb}\pd_\al\Om\pd_\bb\Om,
\end{equation}
where $c$ is the central charge of the undeformed theory.\footnote{In \cite{Aharony:2023dod} the factor $\frac{1}{24\pi}$ was inconsistent with the definition of the energy-momentum tensor used in (1), and should be $\frac{1}{12}$ as in \eqref{eq_weyl_1}.} The formula is derived in appendix \ref{app_weyl}, and the anomaly itself is evaluated there:
\begin{equation}
    e^W = e^{\frac{cI_0}{24} - \frac{\pi c}{6}}\prod_{i=1}^2 \left(\frac{r}{\ep} \right)^\frac{\pi c}{3}.
\end{equation}
In this equation, $\ep$ is the point-splitting regulator, $r$ is some number, and $I_0$ is some finite quantity that is independent of the momentum at large momentum. 

The correlator thus contains the divergent factors:
\begin{equation}
    \prod_{i=1}^2\left[\left(\frac{r}{\ep} \right)^\frac{\pi c}{3}\pr{\frac{|Q_i|}{\ep}}^{2-\Delta} \ep^{\frac{q_i^2}{2\pi|\La|}}\right].
\end{equation}
It will be eliminated by a renormalization of the operators:
\begin{equation}\label{UV_6}
    \OO_i(q_i) \rightarrow \hat{\OO}_i(q_i) \equiv (\bar\mu\ep)^{-2\pi|\La||Q^i|^2} \left(\frac{\ep}{|Q^i|}\right)^{2-\Delta_i}\left( \frac{\ep}{r} \right)^\frac{\pi c}{3} \OO_i(q_i).
\end{equation}
This is the same renormalization of the operators in \cite{Aharony:2023dod}.\footnote{Since $w$ is dimensionless, such are $\ep$ and $\bar\mu$. Therefore, we use $\bar\mu$ to distinguish it from the dimensionful renormalization scale $\mu$ used for the plane.} In particular, it is important that the renormalization does not depend on the modular parameter of the WSt. It does depend on the TSt properties through $\bar\mu$, as will be shown in subsection \ref{subsec_sad_pln}. 

The correlator of the renormalized operators is the remaining finite part:
\begin{equation}\label{eq_dev_cor_1}
\begin{split}
    & e^{|\La|\ATS}\de(q_1+q_2) \int d^2\ttt d\bar{\Om}d\bar{\phi} d^2\sg_1d^2\sg_2\de(\sg_1) \frac{\det(e^{2f_\Om}\tw) J(\bar{e},0) e^{2\bar\Om}\tw^3}{\left|{\det}'\pr{\square}\right|} \cdot 
    \\
    Z_0[g_0]C^0(|\sg_1-\sg_2|) & e^{\frac{cI_0}{24} - \frac{\pi c}{6} + 2\pi in_{1\al}\sg^\al_1 + 2\pi in_{2\al}\sg^\al_2 + \frac{q^2}{2\pi|\La|}\pr{\ln\pr{\frac{|\eta(\ttt)|^6}{\bar\mu^2}} + G_{12}} - |\La|e^{2\bar\Om}\tw - i|\La|\ep^{\al\bb}\ep_{ab}L^a_{\al}\pr{e_c}^b_{\bb} } 
    \\
    = & e^{|\La|\ATS}\de(q_1+q_2) \int d^2\ttt d\bar{\Om}d\bar{\phi} d^2\sg \frac{\det(e^{2f_\Om}\tw) J(\bar{e},0) e^{2\bar\Om}\tw^3}{\left|{\det}'\pr{\square}\right|} \cdot 
    \\
    Z_0[g_0]C^0(|\sg|) & e^{\frac{cI_0}{24} - \frac{\pi c}{6} + 
    2\pi in_{\al}\sg^\al + \frac{q^2}{2\pi|\La|}\pr{\ln\pr{\frac{|\eta(\ttt)|^6}{\bar\mu^2}} + G_{12}} - |\La|e^{2\bar\Om}\tw - i|\La|\ep^{\al\bb}\ep_{ab}L^a_{\al}\pr{e_c}^b_{\bb} }.    
\end{split}
\end{equation}
In the last line we evaluated the $\sg^1$ integral, and denoted $\sg^2,n_{2\al}$ by $\sg,n_\al$.


\newpage
 \section{Saddle-point Evaluation}\label{sec_sad}
In the following section, the high momentum limit of \eqref{eq_dev_cor_1} will be derived. The saddle point approximation (or just ``saddle" from now on) will be used for the remaining integrals. The $\bar\Om,\bar\phi$ integrals will be evaluated in the limit of large TSt-area, compared to the deformation scale. Only two terms in the exponent in \eqref{eq_dev_cor_1} contribute to the saddle:
\begin{equation}\label{eq_exp_sad_OmPhi}
    e^{- |\La|e^{2\bar\Om}\tw - i|\La|\ep^{\al\bb}\ep_{ab}L^a_{\al}\pr{e_c}^b_{\bb} }.
\end{equation}

The $\ttt,\sg_2$ integrals will be evaluated in the large momentum limit, compared to the deformation scale $\frac{q^2}{|\La|}$. The ratio between the momentum $\frac{|q|}{|\La|}$ and the TSt length scale $L$ will dictate the result. The relevant terms in the exponent are:
\begin{equation}
    e^{2\pi in_\al\sg^\al + \frac{q^2}{2\pi|\La|}\pr{\ln\pr{\frac{|\eta(\ttt)|^6}{\bar\mu^2}} + G_{12}}}.
\end{equation}
We will not be able to evaluate the Gaussian integrals around the saddle for the $\ttt,\sg_2$ integrals, and they are subleading in the momentum anyway, so we will not evaluate them at all.

The undeformed correlator $C^0(|\sg|)$ and the partition function $Z_0[g_0]$ are independent of the momentum and, therefore will not contribute to the saddle point approximation. The determinants do depend on the momentum, but in the large momentum limit they go to a constant. This is easily seen for the denominator ${\det}'\pr{\square}$ in \eqref{eq_dev_cor_1}, as $\pd\ln(\bar z) \sim \pd\ln(\pd G)$. This should work in the same way for the numerator (as happened for the plane, (22) in \cite{Aharony:2023dod}).  

\subsection{Constant Conformal-factor and Orientation}\label{subsec_sad_constnat}

The saddle is used with the parameter $l \equiv L\sqrt{|\La|}>>1$, where $L$ is the length scale in $L^a_\al$. Changing variables $x \equiv  e^{\bar\Om}\cos{\bar\phi},\ y \equiv  e^{\bar\Om}\sin{\bar\phi}$ inside the exponent in \eqref{eq_exp_sad_OmPhi} gives:
\begin{equation} 
    \tw\pr{x^2 + y^2} + i\pr{L^1_1\tw + L^2_2 - L^2_1\tn}x + i\pr{L^1_1\tn - L^1_2 + L^2_1\tw}y.
\end{equation}
This function has a saddle point:
\begin{equation}\label{eq_sad_xy}
\begin{split}
    x_s = -\frac{i\pr{L^1_1\tw + L^2_2 - L^2_1\tn}}{2\tw}, \\ 
    y_s = -\frac{i\pr{L^1_1\tn - L^1_2 + L^2_1\tw}}{2\tw}.
\end{split}
\end{equation}
The exponent at this point is:
\begin{equation}
    e^{-|\La|\frac{\pr{L^1_1\tw + L^2_2 - L^2_1\tn}^2 + \pr{L^1_1\tn - L^1_2 + L^2_1\tw}^2}{4\tw}} = e^{-|\La|\ATS \frac{\pr{\tw + T_2}^2 + \pr{\tn - T_1}^2}{4T_2\tw}}.
\end{equation}
where we wrote $L^a_\al$ as in \eqref{eq_vielbein_TSt}, and $\ATS \equiv T_2\pr{X^2 + Y^2} = L^2T_2$ is the TSt area. 

The measure satisfies $e^{2\bar\Om}d\bar\Om d\bar\phi = dxdy$, and the Gaussian integral around the saddle gives $\frac{\pi}{|\La|\tw}$. \eqref{eq_dev_cor_1} is then modified to:
\begin{equation}\label{eq_dev_2}
\begin{split}
    \frac{\pi e^{|\La|\ATS}}{|\La|}\de(q_1+q_2) \int d^2\ttt d^2\sg \frac{\det(e^{2f_\Om}\tw) J(\bar{e},0)\tw^2}{\left|{\det}'\pr{\square}\right|} Z_0[g_0]C^0(|\sg|) \cdot
    \\
    e^{\frac{cI_0}{24} - \frac{\pi c}{6}} e^{2\pi in_{\al}\sg^\al + \frac{q^2}{2\pi|\La|}\pr{\ln\pr{\frac{|\eta(\ttt)|^6}{\bar\mu^2}} + G_{12}} -|\La|\ATS \frac{\pr{\tw + T_2}^2 + \pr{\tn - T_1}^2}{4T_2\tw} }. 
\end{split}
\end{equation}

\subsection{Modular-parameter and Coordinate}\label{subsec_mod_cor}

It remains to perform the integrals over $\ttt$ and $\sg$. Which integral will be performed first depends on the relation between the different scales involved in the correlator. The integral is:
\begin{equation}\label{eq_dev_3}
    \int d^2\ttt d^2\sg e^{2\pi in_{\al}\sg^\al + \frac{q^2}{2\pi|\La|}\pr{\ln\pr{\frac{|\eta(\ttt)|^6}{\bar\mu^2}} + G_{12}} -|\La|\ATS \frac{\pr{\tw + T_2}^2 + \pr{\tn - T_1}^2}{4T_2\tw}}.
\end{equation}
Using $l$ defined in the last subsection and $n \sim |q|L$, the three terms in the exponent can be written schematically as:
\begin{equation}\label{eq_sch_ts_1}
    e^{in\sg + \frac{n^2}{l^2}f_1(\ttt,\sg) + l^2f_2(\ttt)}.
\end{equation}
The large TS-area we used means $l^2 >> 1$. The ratio $r \equiv n/l^2$ dictates the relative coefficients of the three terms of \eqref{eq_sch_ts_1}:
\begin{equation}\label{eq_sch_ts_2}
    e^{in\sg + nrf_1(\ttt,\sg) + \frac{n}{r}f_2(\ttt)}.
\end{equation}
There will be different results for $r<<1$ and for $r>>1$. The first region is where the TS-area is much bigger than the momentum scale. Then, the plane's result is expected to be recovered. The second region is where the momentum is much bigger than all scales in the problem, and it will give another result. The intermediate region is hard to analyze analytically, and there should be some interpolation between these two limits. We will evaluate it numerically for a simple TSt.

\subsubsection{Plane's Limit}\label{subsec_sad_pln}

In the $r<<1$ limit, $\ttt$ is integrated first, considering only the last term of \eqref{eq_sch_ts_2}. The saddle satisfies:
\begin{align}
    \frac{\pd}{\pd\tn}:\ \frac{\tn - T_1}{\tw} = 0, 
    \\
    \frac{\pd}{\pd\tw}:\ \frac{\tw^2 - T_2^2 - \pr{\tn - T_1}^2}{\tw^2} = 0.
\end{align}
It has the solution $\tn = T_1, \tw = T_2$, i.e. $\ttt = T$, and therefore $x=-iX, y=-iY$.\footnote{$x,y$ are imaginary because of the analytic continuation done for $\Om$.}

The exponent at the saddle is:
\begin{equation}
    e^{-|\La|\frac{X^2 + Y^2}{T_2}} =  e^{-|\La|\ATS}, 
\end{equation}
and it cancels an opposite factor in \eqref{eq_dev_2}. The integral over $\sg$ is: 
\begin{equation}\label{eq_sg_integral_1}
    \int d^2\sg e^{2\pi in_\al\sg^\al + \frac{q^2}{2\pi|\La|}\pr{\ln\pr{\frac{|\eta(T)|^6}{\bar\mu^2}} + G(w,T)} }.
\end{equation}
We expect to recover the plane's result, where the saddle behaved as $\sg \sim \frac{q}{|\La|}$ (see paragraph after (109) in \cite{Aharony:2023dod}). Using the parameters $n,r$, it reads $\sg \sim r$. It is then a reasonable guess that the saddle appears for small $\sg$. In particular, we can approximate $G_{12}$ as in \eqref{eq_G_short}, and cancelling the $\ln{|\eta(T)|^6}$ term. Then \eqref{eq_sg_integral_1} becomes:
\begin{equation}\label{eq_sg_integral_2}
    \int d^2\sg e^{2\pi in_\al\sg^\al - \frac{q^2}{2\pi|\La|}\ln{|w|^2}}. 
\end{equation}
We consider the analytic continuation of $|w|^2 = \pr{\sg^1 + T_1\sg^2}^2 + \pr{T_2\sg^2}^2$ as a function of $\sg^1,\sg^2$. The equations for the saddle are:
\begin{align}
    \frac{\pd}{\pd\sg^1}:\ 2\pi in_1 - \frac{q^2}{\pi|\La|}\frac{\sg^1 + T_1\sg^2}{|w|^2} = 0, 
    \\
    \frac{\pd}{\pd\sg^2}:\ 2\pi in_2 - \frac{q^2}{\pi|\La|}\frac{\pr{\sg^1 + T_1\sg^2}T_1 + T_2^2\sg^2}{|w|^2} = 0. 
\end{align}
The solution is\footnote{In the plane, $\sg$ was dimensionful, while here it is dimensionless, which explains the additional $L$ in the saddle. However, this $L$ will be swallowed in the dimensionless regulator $\bar\mu$ to give the mass scale $\mu \equiv \frac{\bar\mu}{L}$, which is the renormalization scale, and has nothing to do with $L$.}:
\begin{equation}
\begin{split}
    |w|^2 &= -\frac{q^2}{\pi^2|\La|^2L^2},
    \\
    \sg^1 + T_1\sg^2 &= -i\frac{q_1}{\pi|\La|L},
    \\
    T_2\sg^2 &= - i\frac{q_2}{\pi|\La|L}.
\end{split}
\end{equation}
Substituting in \eqref{eq_sg_integral_2}:
\begin{equation}\label{eq_cor_plane_limit_1}
    \pr{\frac{\bar\mu|q|}{\pi e|\La|L}}^{-\frac{q^2}{\pi|\La|}}  = \pr{\frac{\bar\mu r}{\pi e}}^{-\frac{nr}{\pi}}.
\end{equation}
This function has a maximum at: $r_{\rm max} = \frac{\pi\sqrt{e}}{\bar\mu}$, which means that in the region of validity of this approximation, the correlator might increase, depending on $\bar\mu$. To clarify this ambiguity, we pay attention to the plane's correlator ((114) in \cite{Aharony:2023dod}):
\begin{equation}\label{eq_cor_plane}
    \pr{\frac{\mu|q|}{\pi e|\La|}}^{-\frac{q^2}{\pi|\La|}}.
\end{equation}
Since any physical result should not depend on the renormalization scale, and $|\La|$ is the only scale in the plane, one should use $\mu = \sqrt{|\La|}$. Then the correlator becomes:
\begin{equation}
    \pr{\frac{|q|}{\pi e\sqrt{|\La|}}}^{-\frac{q^2}{\pi|\La|}}.
\end{equation}
For $\frac{q^2}{|\La|}>>1$, which is the condition for using the saddle approximation in the plane, the function decreases.\footnote{Including the renormalization scale $\mu$, the maximum of \eqref{eq_cor_plane} is at $|q|_{max} = \frac{\pi\sqrt{e}|\La|}{\mu}$. Then for $\sqrt{|\La|}<< |q| < \frac{|\La|}{\mu}$ the correlator might increase. If $\mu = \sqrt{|\La|}$ this region is empty.} Returning to the torus, there are two mass scales $\frac{1}{L},\sqrt{|\La|}$ in the problem, which are relevant for the renormalization scale $\mu \equiv \frac{\bar\mu}{L}$ (and also many others by using the dimensionless quantity $l$).\footnote{Generally, there are two different scales for the two edges of the torus, $L_1 \equiv L$ and $L_2 \sim LT_2$. We consider the generic case where $T_2 \sim 1$. See however } They correspond to $\bar\mu = 1,l$. Since the plane's result is expected to be recovered in this region, the choice $\bar\mu = l$ is preferred\footnote{One can reach the same conclusion by using dimensionful WS coordinates $\tilde\sg = L\sg$, which are now related to the TS coordinates $X$ as in the plane. Then $\frac{\tilde w}{L}$ will appear in $G$, and the singular part at the origin will be $\ln\pr{\frac{\tilde\ep}{L}}$, $\tilde\ep$ being dimensionful as well, the same regulator as in the plane. The renormalization of the operators should be the same as for the plane, and therefore one renormalizes by $(\sqrt{|\La|}\tilde\ep)^{-2\pi|\La||Q^i|^2}$. The factor remaining is $\sqrt{|\La|}L$, which is $\bar\mu$.}, and \eqref{eq_cor_plane_limit_1} reads:
\begin{equation}\label{eq_cor_plane_limit_2}
    \pr{\frac{|q|}{\pi e\sqrt{|\La|}}}^{-\frac{q^2}{\pi|\La|}}  = \pr{\frac{nr}{\pi^2 e^2}}^{-\frac{nr}{2\pi}}.
\end{equation}
The maximum of this function is $\sim \frac{1}{l}$. For using the saddle approximation in this region, one needs $nr>>1$. Hence $\frac{n}{l} >>1 \Rightarrow r>>\frac{1}{l}$, and \eqref{eq_cor_plane_limit_2} decreases.

There is also an additional phase $\pr{-1}^{\frac{q^2}{2\pi|\La|}}$ which can be swallowed in the operators' renormalization. This is the same as (124) in \cite{Aharony:2023dod}. This phase also appears in the $r>>1$ limit and in the special torus considered numerically for all $r$.

\subsubsection{Torus' Limit}

In the $r>>1$ limit, the leading term in the integrand of \eqref{eq_sch_ts_2} depends both on $\ttt$ and $\sg$ and so one can not perform the integrals one after the other as before. Symmetric points are natural candidates for saddle points, and so we pay our attention to the symmetries of $G$.

The $\Th$-function satisfies (see appendix \ref{app_fun}):
\begin{equation}\label{eq_th_prop_w}
\begin{split}
    \Th(-w,\ttt) = -\Th(w,\ttt) & \Rightarrow \Th(\sg^1,\sg^2,\tn,\tw) = \Th(-\sg^1,-\sg^2,\tn,\tw),
    \\
    \overline{\Th(w,\ttt)} = \Th(\bar w,-\bar\ttt) & \Rightarrow \Th(\sg^1,\sg^2,\tn,\tw) = \Th(\sg^1,-\sg^2,-\tn,\tw),
\end{split}
\end{equation}
where $\Th(\sg^1,\sg^2,\tn,\tw) \equiv \Th(\sg^1+(\tn + i\tw)\sg^2,\tn + i\tw)$. \eqref{eq_th_prop_w} implies that the first term of $G$, $-\ln|\Theta(w,\ttt)|^2$ is invariant under\footnote{The last symmetry is a composition of the first two, and in fact, each one of the three can be deduced from the other two. In complex variables, the symmetries are $w \to -w,\ w \to -\bar w \And \ttt \to -\bar\ttt,\ w \to \bar w \And \ttt \to -\bar\ttt$.}:
\begin{equation}\label{eq_th_sym_sg}
\begin{split}
    \sg^1 \to -\sg^1 & \And \sg^2 \to -\sg^2,
    \\
    \sg^1 \to -\sg^1 & \And \tn \to -\tn, 
    \\
    \sg^2 \to -\sg^2 & \And \tn \to -\tn.
\end{split}
\end{equation}
The second term of $G$, $2\pi\tw\pr{\sg^2}^2$, is also invariant under these transformations. Hence, these are symmetries of $G$. These are also symmetries of $|\eta(\ttt)|$.\footnote{$|q| = e^{-2\pi\tw}$ is independent of $\tn$. The other factors are $\pr{1 - q^m}\pr{1 - \bar q^m}$, and they are symmetric with respect to $\tn \to -\tn$, which corresponds to $q \to \bar q$.} The last symmetry can be used to evaluate the saddle in the $\tn,\sg^2$ integrals by assuming that the saddle is at $\tn = \sg^2 = 0$. 

\begin{figure}
\captionsetup{singlelinecheck = false, justification=justified}
\includegraphics[scale=0.4]{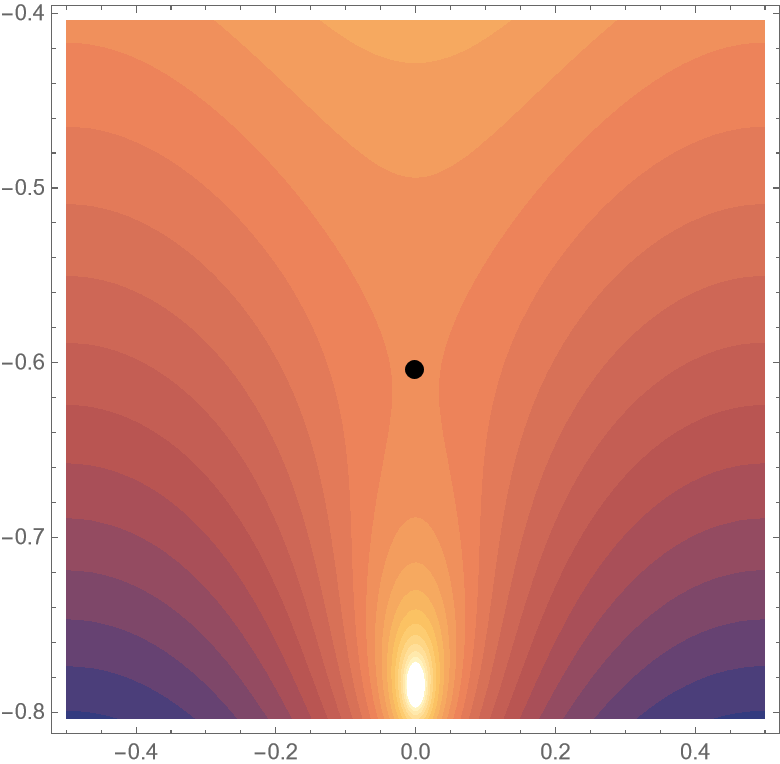}
\quad
\includegraphics[scale=0.4]{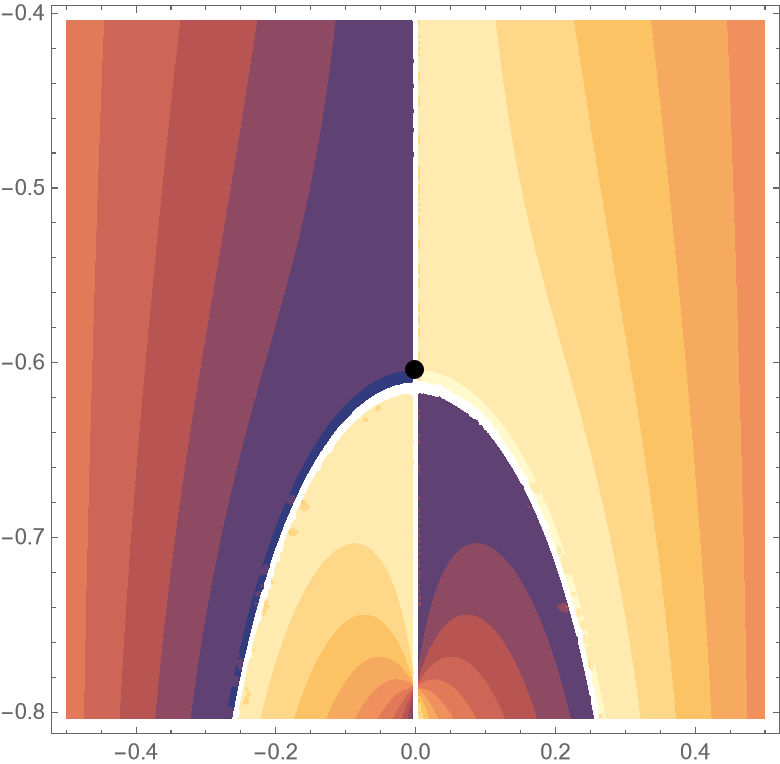}
\caption{Real (left) and imaginary (right) parts of $-\ln\pr{\Th\pr{\sg^1,i\tw}^2}$ around the first saddle at large r ($10$), with $\tw$ given by \eqref{eq_saddle_tw_large_r} ($T_1=0,T_2=10$). The saddle is located at the black point. Two contours of constant phases emerge from the saddle. One of them is the imaginary axis; the other one is near the white line (the brightest color and the darkest color are the same up to $2\pi$), which is the deformed contour. The bright point in the real part is a singularity due to a zero of the $\Theta$-function. The white lines in the imaginary part are discontinuities due to the log.}
\label{fig_large_r_contours}
\end{figure}

One is then left with the $\tw,\sg^1$ integrals. The integral over $\sg^1$ is ill-defined because of the divergence at $\sg^1 = 0$. This divergence results from inserting the two operators in the same place on the WS. It should be avoided due to the point-splitting regularization. To overcome this problem, we note that the function $\Th\pr{\sg^1,i\tw}$ is real, hence $|\Th|^2 = \Th^2$, which is now an analytic function, defined on the ``complex cylinder":
\begin{equation}
    \sg^1 = x + iy,\ x \sim x+1,\ -\infty < y < \infty.
\end{equation}
$x$ is periodic due to the periodicity properties of $\Th$. We define the integration contour in the complex plane, going in a small half circle $\ep e^{i\phi}, \pi \leq \phi \leq 2\pi$ below $\sg^1 = 0$. The branch-cut of the logarithm will be on the positive imaginary axis. The contour is closed, wrapping this cylinder once. We can deform this contour, and the saddle will be where the derivative vanishes:
\begin{equation}\label{eq_sol_tr}
    \frac{d}{d\sg^1} \ln\pr{\Th\pr{\sg^1,i\tw}^2} = 0 \Rightarrow \frac{d\Th\pr{\sg^1,i\tw}}{d\sg^1} = 0.
\end{equation}
This procedure is explained in more detail in appendix \ref{app_com}.\footnote{The same divergence at $\sg=0$ appeared in the plane. The analysis of this case is also given in appendix \ref{app_com}.}
The set of solutions to \eqref{eq_sol_tr} lies on the imaginary axis. On the imaginary axis there are branch points of the log at $y = m\tw$ for any integer $m$, and besides of these $-\ln\pr{\Th\pr{\sg^1,i\tw}^2}$ does not have any other singularities. Between any of these branch points sits one point when the derivative vanishes. Since we do not want the contour to cross them, the first negative one $-\tw<y_s(\tw)<0$ will be the saddle for the $\sg$ integral. The integration contour is deformed appropriately, now going downwards from the saddle down to $y \to -\infty$, see figure \ref{fig_large_r_contours}. We note that along the imaginary axis, the phase of $-\ln\pr{\Th(\sg^1,i\tw)^2}$ is $i\pi$, which will give an overall phase $(-1)^{\frac{q^2}{2\pi|\La|}}$, as for the $r<<1$ limit before.

The saddle's location $y_s(\tw)$ can not be obtained analytically, but numerically, one finds the following simple behavior\footnote{The coefficients were obtained numerically, and the coefficients $\frac{1}{2},1$ appearing in \eqref{eq_y_sad} are $\frac{3.12}{2\pi},\ \frac{6.28}{2\pi}$ respectively. Maybe one can prove that the $\pi$'s are the exact values. In particular, the $-\frac{\tw}{2}$ is seen from a reflection symmetry of the $\Th(iy,\tw)$-function restricted to $(-\tw,0)$ around its center. It is not essential, and we use them for clarity.}:
\begin{equation}\label{eq_y_sad}
    y_s(\tw) \approx
    \begin{cases}
     -\frac{1}{2}\tw - 0.36 \tw^2\ &\tw \lesssim 0.7  \\
     0.17 - \tw \ &\tw \gtrsim 0.7 
     \end{cases}.
\end{equation}

It remains to evaluate the $\tw$ integral. First, we want to verify that it converges. For this aim, we check the small and large $\tw$ domains. Plugging \eqref{eq_y_sad} into $-\ln\pr{\Th(iy_s(\tw),i\tw)^2}$, leads to an almost line\footnote{For $0.7 \lesssim \tw \lesssim 6.5$ one has a slightly different behavior: $-\ln\pr{\Th(iy_s(\tw),i\tw)^2}  \approx 1.95 - 4.73\tw$.  Aside from the linear growth, there is also a logarithmic piece. Hence, as $\tw$ increases, the slope of the line decreases slowly, and the constant decreases rapidly to zero. The slope is approximately obtained by the two first terms in $\Th$: $2q(\tw)^{\frac{1}{8}}\sin(\pi iy_s(\tw))$, giving $-\ln\pr{\Th(iy_s(\tw),i\tw)^2}  \approx \pr{\frac{\pi}{2} - 2\pi}\tw$. The free constant is predicted to be $0.17\cdot 2\pi$, which is not a good approximation due to the other terms.}:
\begin{equation}
    -\ln\pr{\Th(iy_s(\tw),i\tw)^2} \approx 0.05-4.43\tw, \  6.5 \lesssim \tw.
\end{equation}
Hence at large $\tw$ the integrand of \eqref{eq_dev_3} becomes:
\begin{equation}
    e^{\frac{q^2}{2\pi|\La|}\pr{i\pi -\ln\pr{\bar\mu^2}-\frac{\pi\tw}{2} + (0.05 - 4.43 \tw)} -|\La|\ATS\frac{\tw}{4T_2} }.
\end{equation}
It decays fast enough to zero, and the integral converges in this region.

The small $\tw$ region is more difficult to analyze. However, the modular properties of the $\Th,\eta$-functions can be used to extract the leading behavior. We use:
\begin{align}
    \Th\pr{\frac{w}{\ttt},-\frac{1}{\ttt}} & = -i\sqrt{-i\ttt} e^{\frac{i\pi w^2}{\ttt}} \Th(w,\ttt),
    \\
    \eta\pr{-\frac{1}{\ttt}} & = \sqrt{-i\ttt} \eta(\ttt),
\end{align}
and plug $\ttt = i\tw, w = iy$. Then:
\begin{align}
    -\ln\left(\Th\pr{iy,i\tw}\right)^2  & = -\ln\left(\Th\pr{\frac{y}{\tw},\frac{i}{\tw}}\right)^2 + \ln{\tw} - \frac{2\pi y^2}{\tw} + i\pi,
    \\
    \ln|\eta(i\tw)|^6 & = \ln\left|\eta\pr{\frac{i}{\tw}}\right|^6 - 3\ln\tw.
\end{align}
For $1/\tw >> 1$, the r.h.s. can be approximated as\footnote{$\frac{y_s(\tw)}{\tw} \approx -\frac{1}{2}$, therefore  $\sin\pr{\pi \frac{y_s(\tw)}{\tw}} \approx 1$ does not contribute another $\tw^{-1}$ term. This is in contrast to the situation when $\sg^2\neq 0$, see remark \ref{re_large_tau}.},\footnote{The finite part $f_\eta$ for $\eta$ is smaller than $0.006$ for $\tw \lesssim 0.9$, so we did not write it.}:
\begin{align}
    -\ln\pr{\Th\pr{\frac{y_s(\tw)}{\tw},\frac{i}{\tw}}^2} & \approx \frac{\pi}{2\tw} -\ln4,
    \\
    \ln{\left|\eta\pr{\frac{i}{\tw}}\right|}^6 & \approx -\frac{\pi}{2\tw}.
\end{align}
The small $\tw$ limit is therefore:
\begin{align}
    -\ln\pr{\Th\pr{iy_s(\tw),i\tw}^2} & \approx \frac{\pi}{2\tw} + \ln{\tw} -\ln4 - \frac{\pi\tw}{2} + i\pi,
    \\
    \ln|\eta(i\tw)|^6 & \approx -\frac{\pi}{2\tw} - 3\ln\tw.
\end{align}
The integrand of \eqref{eq_dev_3} reads:
\begin{equation}\label{eq_sad_low_tau}
    e^{-\frac{q^2}{2\pi|\La|}\pr{\ln\pr{4\bar\mu^2\tw^2} - i\pi} -|\La|\ATS \frac{|T|^2}{4T_2\tw} },
\end{equation}
for which the integral converges.\footnote{Leaving the $-\ln|\eta(\ttt)|^6$ from $G_{11}$ was crucial for the cancellation of the $\frac{\pi}{2\tw}$.}

As $r$ increases, the maximum of the exponent is reached at small $\tw$, for which the exponent behaves as in \eqref{eq_sad_low_tau}. The saddle is at:
\begin{equation}\label{eq_saddle_tw_large_r}
    \ttt_{2\text{s}} = \frac{\pi|\La|^2\ATS|T|^2}{4T_2q^2} = \frac{\pi|T|^2}{4r^2}.
\end{equation}
Plugging in the integrand yields (ignoring the phase):
\begin{equation}\label{eq_hgh_momen_torus}
    \pr{\frac{2q^2}{\pi e\bar\mu|\La|^2L^2|T|^2}}^{\frac{q^2}{\pi|\La|}} = \pr{\frac{2q^2}{\pi e|\La|^{\frac{5}{2}}L^3|T|^2}}^{\frac{q^2}{\pi|\La|}} = \pr{\frac{2r^2}{\pi el|T|^2}}^{\frac{nr}{\pi}}.
\end{equation}
The result \eqref{eq_hgh_momen_torus} has a minimum at $r_m = \sqrt{\frac{\pi l}{2}}|T|$. We will discuss its meaning in the next section.

One should also consider the other saddle at $\sg^1 = \tn = 0$ and look for the saddle of $\sg^2,\tw$. Since our regularization transforms non-trivially under WS modular transformation, the saddles should not necessarily be the same, and we check which one is larger between the two. We use the identity:
\begin{equation}\label{eq_sad_rel_G}
    G(i\sg^2\tw,i\tw) = \ln(\tw) + G\pr{\sg^2,\frac{i}{\tw}}.
\end{equation} 
The argument $\sg^2$ is real, hence $G\pr{\sg^2,\frac{i}{\tw}} = -\ln\pr{\Th\pr{\sg^2,\frac{i}{\tw}}^2}$. From here the same analysis goes as before, using $y_s(\tw)$ found in \eqref{eq_y_sad}, and the small $\tw$ and large $\tw$ limits of:
\begin{equation}
    \ln|\eta(\tw)|^6 \approx
    \begin{cases}
        -\frac{\pi}{2\tw} - 3\ln(\tw) &\tw <<1 \\
        -\frac{\pi}{2}\tw  &\tw >>1
    \end{cases}
\end{equation}
and
\begin{equation}
    -\ln\pr{\Th\pr{iy_s\pr{\frac{1}{\tw}},\frac{i}{\tw}}^2} \approx 
    \begin{cases}
        0.05-4.43\tw &\tw <<1 \\
        \frac{\pi}{2}\tw - \ln(\tw) - \ln(4) - \frac{\pi}{2\tw} &\tw >>1
    \end{cases}
    .
\end{equation}
At small $\tw$, there might be a saddle resulting from the competition of the logarithm and the power law. But, this saddle is at $\tw \approx \pi$, and so the approximation for small $\tw$ does not hold there. At large $\tw$, there is a competition:
\begin{equation}
    e^{\frac{q^2}{2\pi|\La|}\pr{-\ln\pr{4\bar\mu^2} - \frac{\pi}{2\tw}} - |\La|\ATS\frac{\tw}{4T_2}},
\end{equation}
where the dominant $\tw$ terms are kept. This has a saddle at $\ttt_{2s} = \frac{|q|}{|\La|L} = r$, and the exponent becomes:
\begin{equation}\label{eq_sad_scnd_sad}
    e^{-\frac{q^2}{2\pi|\La|}\ln\pr{4\bar\mu^2} - \frac{|q|L}{4}} \approx (2l)^{-\frac{nr}{\pi}}.
\end{equation}
This saddle decreases with $r$ and is less dominant than the first one we found in \eqref{eq_hgh_momen_torus}, for $r> r_t \equiv \sqrt{\frac{\pi e}{4}}|T| \sim |T|$. Therefore, \eqref{eq_hgh_momen_torus} is the correlator at the very large momentum limit.

\subsection{Intermediate Region}\label{subsec_sad_intreg}

\begin{figure}
\captionsetup{singlelinecheck = false, justification=justified}
\includegraphics[scale=0.5]{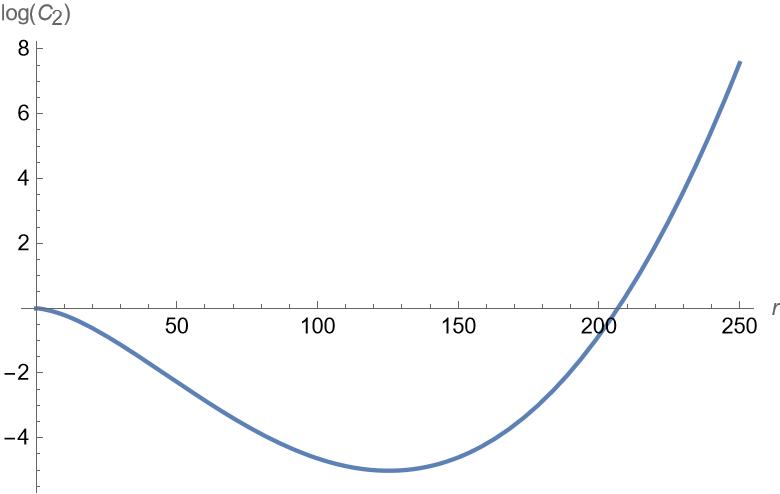}
\qquad
\includegraphics[scale=0.5]{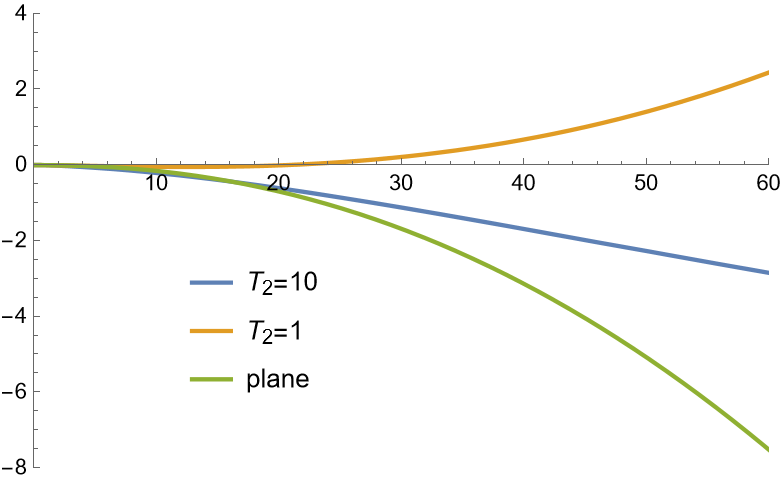}
\caption{Left: Saddle point approximation of the two-point correlator at large momentum as a function of $r \equiv \frac{|q|}{|\La|L}$ for a rectangular torus with $T_2=10$. The minimum is not at $r\approx 1$ since $T_2$ and $l$ are not of order unity. Right: Close up to the small $r$ region, compared to the plane and another rectangular torus with $T_2=1$ (shifted appropriately), for which the minimum is at $r \approx \sqrt{l}=10$.}
\label{fig_cor_torus}
\end{figure}

For the special case of TSt with $T_1=0$ and momentum aligned along one of the torus axes $n_2=0$, one can find the transition between the two regions of small and large $r$. \eqref{eq_dev_3} takes the form:
\begin{equation}\label{eq_sadddle_special_torus}
    \int d^2\ttt d^2\sg e^{in\sg^1 + \frac{nr}{2\pi}\pr{\ln\pr{\frac{|\eta(\ttt)|^6}{\bar\mu^2}} + G(w,\ttt)} - \frac{n}{4r} \frac{\pr{\tw + T_2}^2 + \tn^2}{\tw} },
\end{equation}
where $n \equiv |q|L = 2\pi n_1$. The relations $T_1 = n_2 = 0$ allow to take the saddle $\sg^2 = \tn = 0$ for $n>>1,nr>>1$ and all $r$.\footnote{The condition $n>>1$ is needed for this exponent to be dominant for the saddle approximation. It is also specifically needed for the $r \sim 1$ region. The condition $nr>>1$ is needed for the $r<<1$ region, where the saddle of $\sg^2$ comes from the second term alone.} Then one finds numerically $\sg^1(\tw) = iy_s(\tw)$.\footnote{The relation \eqref{eq_y_sad} was for large $r$.} Finally, the saddle of $\tw$ of the whole expression is found as a function of $r$:
\begin{equation}\label{eq_numerical_saddle}
    e^{-ny_s(\tw) + \frac{nr}{2\pi}\pr{\ln\pr{\frac{|\eta(i\tw)|^6}{\bar\mu^2}} - \ln\pr{\Th(iy_s(\tw),i\tw)^2}} - \frac{n}{4r} \frac{\pr{\tw + T_2}^2}{\tw} }.
\end{equation}
We used the values: 
\begin{equation}
    T_2=10,\ l=\bar\mu=100,\ 0.01 \leq r \leq 250.
\end{equation}
where $T_2=10$ and not $1$, so its transition to $0$ will be clear. $l=100$ such that for the whole range of $r$ the conditions $n=rl^2>>1,nr=(rl)^2>>1$ are satisfied. The final result is shown in figure \ref{fig_cor_torus}. The correlator decreases and afterward increases, as calculated analytically. The minimum depends on $T_2$, as can be seen by comparing $T_2=1$ to $T_2=10$. The plane's correlator is also present for comparison.

\begin{figure}
\captionsetup{singlelinecheck = false, justification=justified}
\includegraphics[scale=0.5]{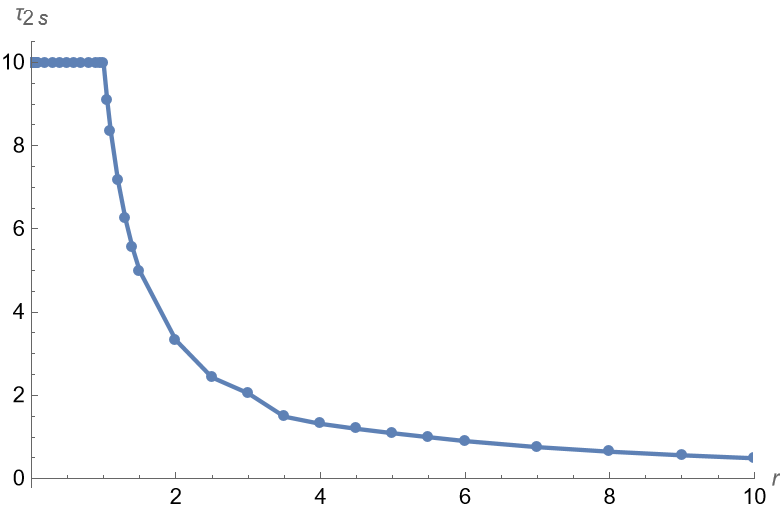}
\quad
\includegraphics[scale=0.5]{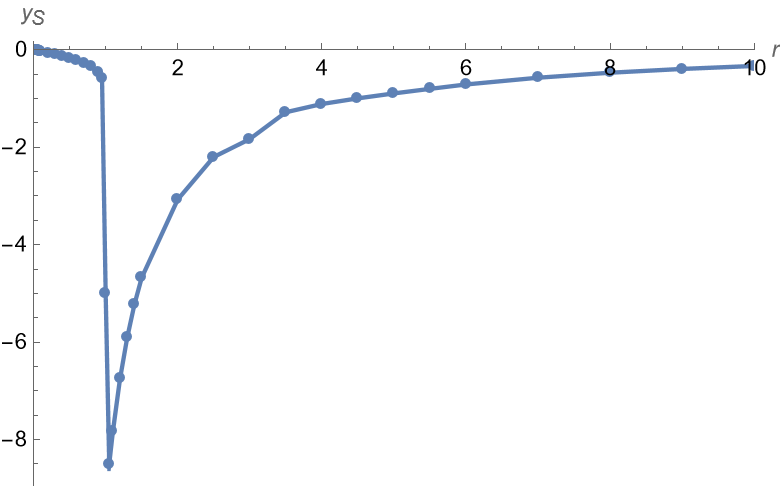}
\caption{$\tw$ (left) and $\sg^1$ (right) saddle point as a function of $r \equiv \frac{|q|}{|\La|L}$. $T_2=10$.}
\label{fig_t2sg1_saddle}
\end{figure}

$\ttt_{2s}(r)$ is shown in the left part of figure \ref{fig_t2sg1_saddle}. At small $r$ the saddle of $\ttt_{2s}(r)$ is the same as $T_2$. At $r=2\pi$, there is a transition because the second term in the exponent \eqref{eq_numerical_saddle} becomes relevant. Eventually, it controls the large $r$ behavior, taking $\ttt_{2s}(r) \sim r^{-2}$ to zero as predicted by \eqref{eq_saddle_tw_large_r}. 

\begin{figure}
\captionsetup{singlelinecheck = false, justification=justified}
\includegraphics[scale=0.5]{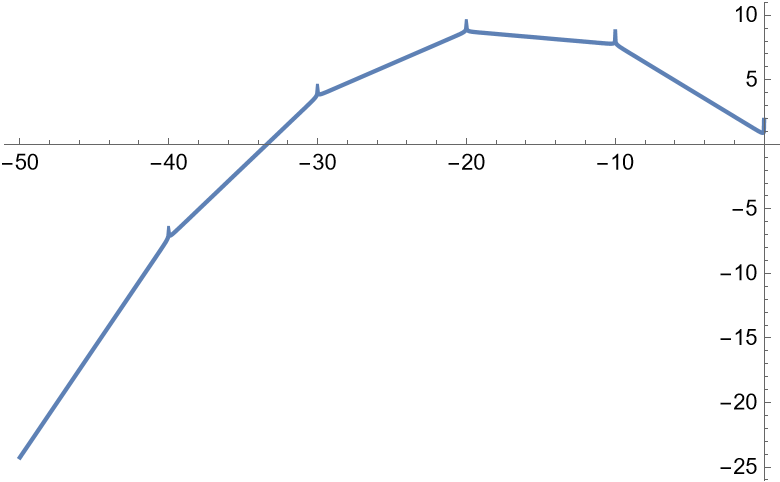}
\qquad
\includegraphics[scale=0.5]{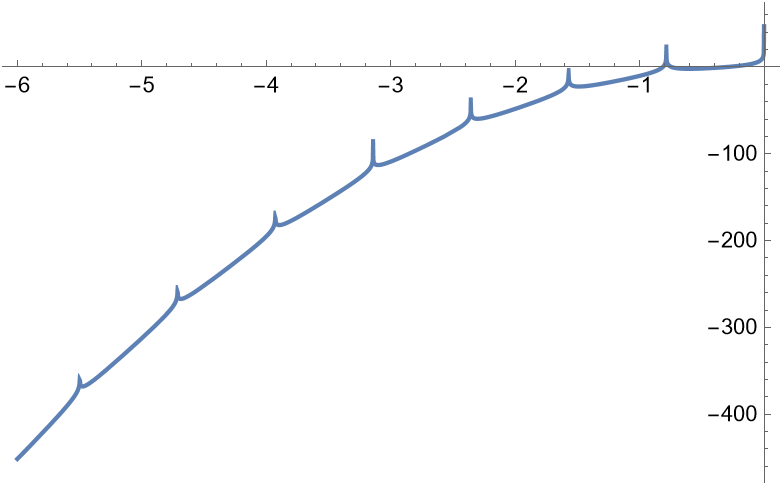}
\caption{Saddles for the torus at small r (left) and large r (right). The saddles are located at the minima. The ``spikes" are the zeros of $\Th$, located at $\ttt_{2s}(r)$, with $\ttt_{2s} = 10,\ \pi/4$ respectively, with $T_2=10$. Note the difference of scales in the $x,y$-axes.}
\label{fig_sad}
\end{figure}

$\sg^1_s(\ttt_{2s}(r))$ is shown in the right part of figure \ref{fig_t2sg1_saddle}. At small $r$, the absolute value of the saddle increases linearly with $r$. After the transition point, it decreases with $\tw$, see \eqref{eq_y_sad}. The sharp transition can be understood as follows. Along the imaginary axis, there are many singularities and extremum points between them due to the $-\ln\Th^2$ term. The extremum points are right to the singularities or in the middle between them; see the right part of figure \ref{fig_sad}. However, for small $r$, the first are ``plane's like", in the sense that the $-2\pi ny$ term (which at large enough $y$ is negligible anyway) changes their location, now being left to the singularity; see the left part of figure \ref{fig_sad}. In particular, the first extremum point, which is picked by the $\sg^1$ contour integral, is exactly the same as for the plane $y \approx -\frac{r}{\pi}$. As $r$ increases, the ``plane's like" saddles disappear gradually, where at $r \approx 1$, even the first negative extremum is dictated by $-nr\ln\Th^2$ alone. From there on, it decreases with $\ttt_{2s}(r)$. See appendix \ref{app_com} for more details.

\newpage
 \section{Summary and Discussion}\label{sec_sum}
In this section, we summarize the main steps of the calculation done in the last two sections, discuss the physical interpretation of the result \eqref{eq_hgh_momen_torus}, and compare it with other non-local field theories.

\subsection{Calculations' Summary}

The calculation of $\TT$-deformed correlation functions in the JT-formalism contains the following steps. First, one has to remove the diffeomorphism invariance. This is done by choosing a gauge, where for $\TT$-deformed CFTs, the orthogonal gauge $\bar e = e^{\Om+\ep\phi}M$ turns out to be convenient since it makes the undeformed fields' PI trivial. The FP-determinant might be hard to calculate; however, it is unnecessary for the large momentum limit.

Second, one has to deal with the various PIs. They are simple, but various subtleties arise:
\begin{itemize}
    \item  The vielbein $e$ PI is trivial due to the orthogonal gauge chosen for the FP-constraint $\de(e-\bar e)$.\footnote{One could use other gauges, for example the static gauge $X \sim \sg$. Then, the $X$ PI would be trivial, but others will be complicated. We managed to solve all of these problems in this gauge.} 
    
    \item The TS-coordinates $X$ dependence is linear in an exponent, and the PI gives a $\de$-function $\de(K)$. The only complication here is that the coefficient is not purely imaginary, and an analytic continuation should be done to the conformal factor $\Om$. 
    
    \item The conformal factor $\Om$ and orientation $\phi$ PIs are trivial due to the constraint $\de(K)$. The complication here is that $K=0$ does not fix the solution $\Om_0,\phi_0$ completely; the kernel of its linearized from $Q$ is not trivial and leads to a two-parameter set of solutions, labeled by $\bar\Om,\bar\phi$. Integration over them is left from the whole $\Om,\phi$ PI, and one needs to derive its measure using the properties of $Q$.
    
    \item The remaining PI is that of the undeformed fields $\psi$, which is trivial when the undeformed theory is a CFT and the orthogonal gauge is chosen. The complication here is to derive the Weyl anomaly properly.
\end{itemize}

Then, one is left with some finite number of integrals of a divergent expression, which should be regularized. The divergences appeared in three places, all related to the insertion of local operators $O(\sg_i)$ in the undeformed theory (in the metric at the operators' positions $g(\sg_i)$, in the WS area, and in the Weyl anomaly). They make the metric diverge, much like a charged point particle causes the electric field to diverge. In this spirit, one natural way to regulate it is by changing the point-like particle to a small spherical shell. The analog for our calculation is point-splitting regularization, which is also preferred since it maintains the TS translation symmetries. It also deals with another potential divergence when two (or more) operators coincide.

The final integrals include the positions of all inserted operators $\sg_i$, the two modes $\bar\Om,\bar\phi$, and all moduli of the WS, which is $\ttt$ for the torus and none for the plane. These integrals can not be solved exactly, but in the large momentum limit, they can be solved using the saddle point approximation, exploiting various symmetries that exist in the problem. 

Some integrals need an analytic continuation to the complex plane, where the integration contour should be deformed carefully according to the singularities of the integrated. For the plane this procedure was not complicated, compared to the torus, where the order of integration as well as the contour deformation depends on $r$.

\subsection{Physical interpretation}

For interpreting the large momentum correlator, it is beneficial to compare it with the planes' result. In the plane, the deformation scale $t$ is the only scale in the deformed theory. In contrast, in the torus, an additional scale appears: the torus' length scale $L$. The relation between the two scales is a dimensionless quantity $l^2 \equiv L^2/t$, which can be compared to the dimensionless momentum scale $n \sim |q|L$. In this terminology, high momentum in the plane means $tq^2 = \frac{n^2}{l^2} = nr >> 1$, where $r \sim \frac{t|q|}{L}$ as before is the ratio between the length scale associated with the smeared operators and the torus' length scale. Then \eqref{eq_int_pln} reads $\pr{\frac{nr}{\pi e}}^{-\frac{nr}{2\pi}}$. This is  $r<<1$ torus' result \eqref{eq_cor_plane_limit_2}, where $L$ is the largest scale in the problem (the plane's limit is $L\to\infty$). In fact, some ambiguities in the whole calculation were chosen to achieve this equality. 

The new result, where the torus characteristics are manifest, is for $r>>1$. The correlator \eqref{eq_hgh_momen_torus} has two interesting properties:
\begin{itemize}
    \item At this region of $r>>1$, it increases with the momentum, unlike for the infinite plane and the large torus' limit $r<<1$, where the correlator decreases.

    \item It depends on an IR quantities $L,|T|$, so it presents a UV-IR mixing. 
\end{itemize}
The first property, together with insights from other non-local theories (see below), suggests the following interpretation. The deformed operators are smeared over a length scale $t|q|$. For $r<<1$, this length scale is much smaller than $L$, and the correlator decreases, as for the plane.\footnote{But both the area and the momentum are large; namely, the following hierarchy $1<<l<<n<<l^2$ was considered.} This happens without any disruption for the infinite plane. However, for the finite TSt, at large enough momentum, the operators cover the whole torus, and the correlator need not decay from that point on. In particular, \eqref{eq_hgh_momen_torus} shows that the correlator grows. The transition between these two $r$ limits was hard to calculate analytically, and we found it numerically in the special case of a rectangular torus and momentum aligned along one of its axes. The result was shown in figure \ref{fig_cor_torus}, decreasing at small $r$ and increasing at large $r$, supporting this argument. 

However, it seems that the situation is a little bit more complicated than this schematic picture. Two additional scales appear when considering the two saddles found for the large $r$ region, given is \eqref{eq_hgh_momen_torus} and \eqref{eq_sad_scnd_sad}. The second saddle \eqref{eq_sad_scnd_sad} is dominant as long as $|q|t\lesssim L|T|$, which might hint that the scale $L|T|$ plays a role here. This scale is the other edge of the TSt, which is different from the first one when $T_2>>1$. In addition, \eqref{eq_hgh_momen_torus} by itself hints to another scale; the saddle decreases until the momentum satisfies $|q|t = \sqrt{\frac{\pi l}{2}}L|T|$. It suggests another scale, which is related but not the same as $L|T|$. Furthermore, even if $|T| \sim 1$, it is not the same as $L$ since $l>>1$. Only after this scale the correlator increases. 

\begin{figure}
\captionsetup{singlelinecheck = false, justification=justified}
\includegraphics[scale=0.5]{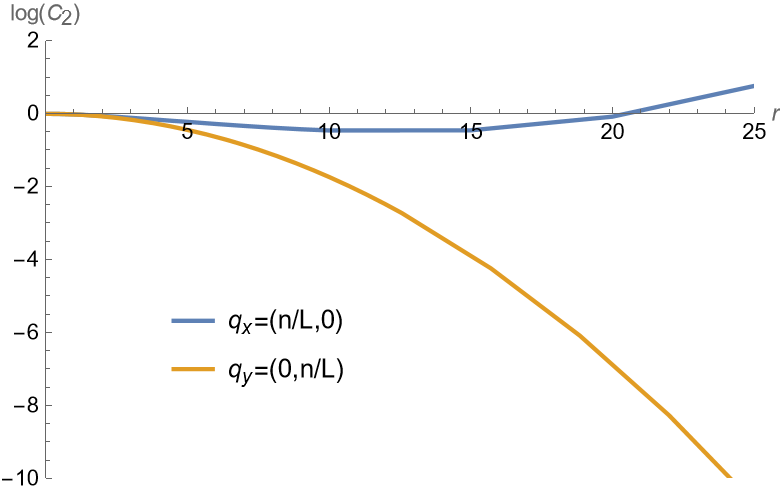}
\qquad
\includegraphics[scale=0.5]{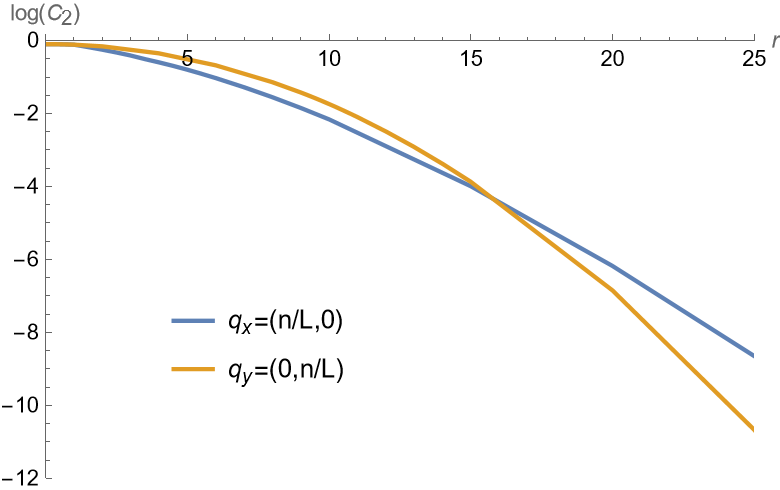}
\caption{Saddles for the torus for momenta aligned along the two edges for $T_2=1$ (left) and $T_2=10$ (right). The crossing point exists for $T_2=10$ at $r_t \approx 15 \approx \sqrt{\frac{\pi e}{2}}T_2$}.
\label{fig_com_momenta}
\end{figure}

One natural question to ask about this suggestion regards the orientation of the smeared operator, if exists at all. For example, let us assume that the operator is smeared along the momentum's direction. One considers a rectangular TSt with $T_1=0$ and a separation of scales between its edges $T_2>>1$. Then, the result might depend on its orientation for momentum in the intermediate region $L<|q|t<LT_2$. If $q$ is aligned along $L$, then the correlator should increase with $q$ as it already wraps this edge. If $q$ is aligned along $LT_2$, then the correlator should decrease with $q$ until $|q|t \sim LT_2$. The analytical results presented before at small and large $r$, \eqref{eq_cor_plane_limit_2} and \eqref{eq_hgh_momen_torus} correspondingly, which are for generic TSt, depend on $q^2$, which means that the orientation of the momentum does not play any role in these two regions, as expected. The saddle for $q_x=\pr{\frac{n}{L},0}$ was presented at subsection \ref{subsec_sad_intreg}. Whether this is the leading saddle or \eqref{eq_hgh_momen_torus} in the intermediate region $1\lesssim r \lesssim LT_2$, the correlator decreases. Hence, the correlator ``knows'' about the other edge and does not increase from $r \approx 1$. One can also compare the correlator for saddles obtained for momentum aligned along each edge. The saddle for $q_y=\pr{0,\frac{n}{L}}$ is obtained from the analog of \ref{eq_sadddle_special_torus}
\begin{equation}
    \int d^2\ttt d^2\sg e^{inT_2\sg^2 + \frac{nr}{2\pi}\pr{\ln\pr{\frac{|\eta(\ttt)|^6}{\bar\mu^2}} + G(w,\ttt)} - \frac{n}{4r} \frac{\pr{\tw + T_2}^2 + \tn^2}{\tw} }.
\end{equation}
Due to the symmetries of the integrand, one takes $\sg^1 = \tn = 0$. Then one uses \eqref{eq_sad_rel_G} and analytic continue $\sg^2$ as before, seeking the saddle on the imaginary axis $\sg^2 = iy_s(\tw)$:
\begin{equation}
    e^{-nT_2y_s(\tw) + \frac{nr}{2\pi}\pr{\ln\pr{\frac{|\eta(i\tw)|^6}{\bar\mu^2}} + \ln(\tw) - \ln\pr{\Th(iy_s(\tw),\frac{i}{\tw})^2}} - \frac{n}{4r} \frac{\pr{\tw + T_2}^2}{\tw} }.
\end{equation}
The large $r$ saddle will be the same as \eqref{eq_sad_scnd_sad}, and so one should trust this result at most until it reaches $r_t$. The two saddles are presented on the right side of figure \ref{fig_com_momenta}. Both present the same plane behavior at small $r$. Then both decrease, and the saddle for $q_y$ is higher until the crossing point is $r_t$. Since the correlator is continuous, and these two saddles give the correct correlator for these two momenta at small $r$, it seems that they describe the correlator for each momenta up to $r_t$, and from that point on the saddle of $q_x$ is relevant for both. Therefore, we do see a difference between different directions, but both decrease until the smearing scale reaches the lengthier edge. This issue requires further work. For comparison, on the left side of figure \ref{fig_com_momenta}, the correlators for $T_2=1$ are present. In this case, the two edges are symmetric, and no difference is expected between $q_x$ and $q_y$. Indeed, they both start as the plane, and then the saddle of $q_x$ is relevant for both, without any transition point. 

One comment is in place. The $\TT$-deformation is defined for a theory living in a two-dimensional flat space. In addition to the plane and the torus, one can consider another simple geometry: the cylinder. This geometry also has two translation symmetries for the TS; therefore, one can define the Fourier transform of position space operators. The correlation between the Fourier-transformed operators can be calculated similarly to what was done for the plane and the torus. It would be interesting to see the behavior of the correlator at large momentum, for momentum aligned along the compact and the non-compact directions. According to the reasoning given here, it is expected to present an exponential decay, as for the plane, not depending on the orientation of the momentum.

\subsection{Comparison to Other Non-Local Field Theories}

The physical interpretation presented in the last subsection was inspired by properties similar to another type of non-local field theory, non-commutative gauge theories \cite{Gross:2000ba, Rozali:2000np}. One main similarity between non-commutative gauge theories and $\TT$-deformed theories is that operators are well-defined in momentum space. In particular, they share some more specific features: momentum-dependent renormalization, independence of the initial IR data, UV-IR mixing, and exponential dependence on the momentum. Therefore, the same interpretation for the latter was used here, namely that operators are smeared over a length scale $t|q|$, which explained the behavior of their correlations in the plane and in the torus. However, in non-commutative gauge theories,  operators are smeared in a direction orthogonal to the momentum; here, they are smeared uniformly in all directions. 

Two more types of non-local field theories present well-defined operators in momentum space, and their correlators at large momentum behave schematically as momentum to the power of momentum. These theories, as well as the non-commutative gauge theories, are different in their structure from the $\TT$-deformed theories discussed here, so there is no reason to expect their correlation functions to agree with what we calculated, and indeed, they present different correlators. The first type is Dipole theories \cite{Bergman:2001rw}, where Non-commutative gauge theories themselves can be considered as dipole theories. Generally, they are non-local but do not present a UV-IR mixing. There, also a non-locality scale emerges, which is proportional to the dipole's length. The correlator also distinguishes the dipole's direction from other directions, unlike the uniform dependence presented here at high momentum.

The second type appears in string theory in holographic descriptions of an irrelevant deformation \cite{Asrat:2017tzd}. This deformation resembles the $\TT$-deformation in some way; it is the ``single-trace'' version of the $\TT$-deformation when applied to symmetric orbifold CFTs. Their UV limit is the same as that of Little String Theory (they have the same bulk description there). The latter has no local operators  but only operators in momentum space (it is invariant under T-duality for instance). The correlators there also do not have any preferred direction.

\section*{Acknowledgments}

The author would like to thank Ofer Aharony for general guidance during the work and many useful discussions. This work was supported in part by an Israel Science Foundation (ISF) grant number 2159/22, by Simons Foundation grant 994296 (Simons Collaboration on Confinement and QCD Strings), by the Minerva Foundation with funding from the Federal German Ministry for Education and Research, by the German Research Foundation through a German-Israeli Project Cooperation (DIP) grant “Holography and the Swampland”, and by a research grant from Martin Eisenstein.


\appendix

\newpage
\section{Modular Transformations}\label{app_mod}
In this appendix, we briefly describe the modular transformation. We refer the reader also to \cite{DiFrancesco:1997nk}, pp.338-340. Here, we also add comments relevant to this work. The modular transformations are transformations of the modular parameter $\ttt$, having the structure of the $PSL(2,\mathbb{Z})$ group: 
\begin{equation}
    \ttt \to \frac{a\ttt + b}{c\ttt + d}.
\end{equation}
$a,b,c,d \in \mathbb{Z}$ and $ad - bc = 1$, since we keep the orientation of the torus. It is trivial that $\mathcal{R}:\ttt \to \bar \ttt$ represents the same torus. Hence, each orientation preserving transformation $\mathcal{P}$ is accompanied by a changing orientation transformation $\mathcal{RP}$ (for which $ad - bc = -1$). The $PSL(2,\mathbb{Z})$ group is generated by:
\begin{align}
    \mathcal{T}:& \ttt \to \ttt + 1,
    \\
    \mathcal{S}:& \ttt \to -\frac{1}{\ttt}.
\end{align}

The modular parameter $\ttt$ does not contain the whole information about the torus - the length of one edge is missing. In general, a torus is described by two independent vectors on a lattice. They can be described as two numbers in the complex plane  $\om_{1,2}$. Then the modular parameter $\tau \equiv \frac{\om_2}{\om_1}$ cares only about the angle between them and the ratio between their lengths. Therefore, it is enough when parameterizing a torus for a CFT. However, for a general QFT, one needs $\om_{1,2}$. In terms of these, the modular transformations reads\footnote{There is a mistake in Eq.(10.6) of \cite{DiFrancesco:1997nk} according to its definition of $\ttt$ in pp.337.}:
\begin{equation}
    \begin{pmatrix} \om_2 \\ \om_1 \end{pmatrix} 
    \to
    \begin{pmatrix}
    a & b \\ c & d 
    \end{pmatrix} 
    \begin{pmatrix} \om_2 \\ \om_1 \end{pmatrix}. 
\end{equation}
The new vectors are another two independent vectors, as the matrix is invertible; they are two lattice vectors as the matrix elements are integers; and it preserves the area of the torus since the determinant of the matrix is unity. The matrices for the two generators are:
\begin{align}
    \mathcal{T}: 
    \begin{pmatrix}
    1 & 1 \\ 0 & 1    
    \end{pmatrix}
    ,\ 
    \mathcal{S}: 
    \begin{pmatrix}
    0 & -1 \\ 1 & 0    
    \end{pmatrix}.
\end{align}
Without loss of generality, we take $\om_1 = L$ to be on the real axis. When applying a modular transformation, $L$ should be modified accordingly, so the torus' area $L^2\tw$ remains the same. For example, $L \to L|\ttt|$ for $\mathcal{S}$. 

Using two dimensional vectors rather than complex numbers, one has:
\begin{equation}\label{eq_mod_L}
    \begin{pmatrix} L_2 \\ L_1 \end{pmatrix} \to \begin{pmatrix} a & b \\ c & d \end{pmatrix} \begin{pmatrix} L_2 \\ L_1 \end{pmatrix}.
\end{equation}
 The choice $\om_1 = L$ reads:
\begin{equation}
    L_1^a = L
    \begin{pmatrix} 
    1 \\ 0 
    \end{pmatrix},\ 
    L_2^a = L
    \begin{pmatrix} 
    \tn \\ \tw 
    \end{pmatrix}.    
\end{equation}
The vielbein \eqref{eq_vilebein_sg} and $L^a_\al$ of the TSt \eqref{eq_vielbein_TSt} are essentially these two, without putting $\om_1$ on the real axis, but allow it to be at any direction, and $L = e^\Om$ for the WSt. Its direction is not important for the fixed TSt but must be included for the WSt, where all angles at all $\sg$ are integrated.

If considering integration over the modular parameter only, the measure $\frac{d^2\ttt}{\tw^2}$ is modular invariant. It is trivial for $\mathcal{T}$, and for $\mathcal{S}$ one has $d^2\ttt' = \frac{d^2\ttt}{|\ttt|^4}$.\footnote{This form resembles the measure $\frac{dx}{x}$ which is invariant under $x \to 1/x$ transformation. The $|\ttt|^{-4}$ is also expected since $\ttt' \sim \frac{1}{\ttt} \Rightarrow \frac{d\ttt'}{d\ttt} \sim \frac{1}{\ttt^2} \Rightarrow \det{\frac{d\ttt'}{d\ttt}} \sim \frac{1}{\ttt^4}$.} If considering integration over all possible tori, then one can integrate over all $L^a_\al$ with $d^4L$ which is modular invariant.\footnote{Using \eqref{eq_mod_L}, the Jacobian for the four components $L^a_\al$ reads $(ad - bc)^2 = 1$.} Alternatively one can use $d\Om e^{2\Om} d\phi \frac{d^2\ttt}{\tw}$, which is also modular invariant.\footnote{It equals $Ad\Om  d\phi \frac{d^2\ttt}{\tw^2}$ with $A = e^{2\Om}\tw$ the torus' area. Any other power of $A$ is equally good, e.g. $d^4L = e^{4\Om}d\Om d\phi d^2\ttt = A^2 d\Om d\phi \frac{d^2\ttt}{\tw^2}$. These two measures appeared in \cite{Dubovsky:2018bmo}, see (43) and (44).}

In our theory, we have two tori: the TSt and the WSt. The correlator, which depends on the TSt, should not depend on the vectors we choose to describe the torus, hence it is expected to be invariant under $PSL(2,\mathbb{Z})_{TS}$. It also implies that we can two representative vectors that are convenient for us. Indeed this is a global symmetry of the action \eqref{eq_per_action2}. This can be seen as follows. $S_0$ does not depend on the TSt, and so we need to check $S_{JT}$ alone. We denote $E^a_\al \equiv \pd_\al X^a - e^a_\al$.  Then, using matrix notation, $S_{JT}$ can be written as:
\begin{equation}
    S_{JT} = \int d^2\sg\Tr(E\ep E^T\ep^T).
\end{equation}
Modular transformations of the TSt acts as $E^a_\al \to A^a_bE^b_\al$, with $A\in PSL(2,\mathbb{Z})$. It changes the allowed domain for $X^a$ and the two TS vectors $e^a_{1,2}$. The Lagrangian changes as $\Tr(E\ep E^T\ep^T) \to \Tr(AE\ep E^T A^T\ep^T) = \Tr(E\ep E^T A^T\ep^T A)$. Now $A^T\ep^T A = \ep^T$, since it is just the determinant of $A$, which is unity.

Modular transformations of the WSt are not global symmetries. Since we integrate over all possible WS tori, they are part of the diffeomorphism invariance of the theory. When considering all diffeomorphisms preserving the coordinates' domain \eqref{eq_sg_domain}, they correspond to the large diffeomorphisms:
\begin{equation}\label{eq_modular_sg}
    \begin{pmatrix} \sg^1 \\ \sg^2 \end{pmatrix} 
    \to
    \begin{pmatrix}
    a & -b \\ -c & d 
    \end{pmatrix} 
    \begin{pmatrix} \sg^1 \\ \sg^2 \end{pmatrix}. 
\end{equation}
The gauge-fixing of the diffeomorphism invariance done in subsection \ref{subsec_gauge} only deals with small diffeomorphisms. The large diffeomorphisms $PSL(2,\mathbb{Z})_{WS}$ remaining appear now as modular transformations. The two metrics with $\ttt' = \frac{a\ttt + b}{c\ttt + d},\ e^{2\Om'} = e^{2\Om}|c\ttt + d|^2$ describe the same WSt, as they are related by the change of coordinates \eqref{eq_modular_sg}. They are residual gauge transformations. One usually eliminates them by reducing the integration domain of $\ttt$ from the upper half plane to some fundamental region.

However, in our theory, the situation is slightly different. We integrate over all possible mappings $X^a(\sg)$ from the WSt to the TSt. The relation \eqref{rel_X_Y} sends specific WSt cycles $C_{WS}$ to specific TSt cycles $C_{TS}$, and so it is a global gauge-fixing. Large diffeomorphisms do not respect this relation. From another perspective, the transformation $\ttt \to \ttt', \Om \to \Om'$ described before, although describing the same WSt, changes $C_{WS}$. According to this, one has two options:
\begin{itemize}
    \item The first is to fix some cycles of the TSt $C_{TS}$, which means fixing  $L^a_\al$. The modular parameter $\tau$ is then integrated over the upper half plane.

    \item The second is to integrate $\ttt$ only in the fundamental domain, which means we fix some WSt $C_{WS}$. Then, one needs to sum over different $L^a_\al$ related by $SL(2,\mathbb{Z})_{TS}$ transformations.
\end{itemize}
In this work, we chose the first option. We fixed some $L^a_\al$ (with $T$ in the fundamental region) and integrated $\ttt$ over the whole upper half plane. We note that now the $PSL(2,\mathbb{Z})_{TS}$ ``combines" with $PSL(2,\mathbb{Z})_{WS}$, i.e. applying the same modular transformation to both tori is still a symmetry. This is because the same WS cycles are mapped to the same TS cycles. $PSL(2,\mathbb{Z})_{TS}$ transformation alone is not a symmetry anymore since it does not preserve the gauge-fixing condition \eqref{rel_X_Y}. It is broken in the quantum theory; the regularization we choose breaks these combined $PSL(2,\mathbb{Z})$ transformations. Since $PSL(2,\mathbb{Z})_{WS}$ is part of the diffeomorphism invariance of the theory, it can not be broken, and therefore, the global $PSL(2,\mathbb{Z})_{TS}$ is broken.


\newpage
\section{Determinants}\label{app_det}
\subsection{P}

The FP-determinant $J$, appearing in \eqref{eq_FP_identity}, is related to the operator $P$, which is the linearization of the constraints:
\begin{equation}
\begin{split}
    e^{(V)} - \bar e(\Om(\sg),\phi(\sg),\ttt) &= 0
    \\
    \sg_1^{(V)} - 0 &= 0
\end{split}
\end{equation}
around their solution $\bar e,0$. The operator $P$ is given by:
\begin{equation}\label{P_def_one}
    P
    \begin{pmatrix}
    \de V^{\bb} \\ \de\Om \\ \de\phi \\ d\ttt_i
    \end{pmatrix}
    =
    \begin{pmatrix}
        \pd_{\bb}\bar{e}^a_{\al} + \bar{e}^a_{\bb}\pd_{\al} 
        & 
        -\bar{e}^a_{\al}
        &
        -\ep^a_b\bar{e}^b_{\al}
        &
        -\pd_{\ttt_i}\bar{e}^a_\al
        \\
        \de^\al_\bb \int d^{2}\sg\de(\sg) 
        & 0 & 0 & 0
    \end{pmatrix}
    \begin{pmatrix}
    \de V^{\bb} \\ \de\Om \\ \de\phi \\ \de\ttt_i
    \end{pmatrix},
\end{equation}
where
\begin{equation}
\begin{split}
    \pd_{\bb}\bar{e}^a_{\al} & = \bar{e}^a_{\al}\pd_{\bb}\Om + \ep^a_b\bar{e}^b_{\al}\pd_{\bb}\phi, 
    \\
    \pd_{\tn}\bar{e}^a_\al & = e^\Om
    \begin{pmatrix}
        0 & \cos(\phi) \\ 0 & -\sin(\phi)    
    \end{pmatrix},
    \\
    \pd_{\tw}\bar{e}^a_\al & = e^\Om
    \begin{pmatrix}
        0 & \sin(\phi) \\ 0 & \cos(\phi)    
    \end{pmatrix}.
\end{split}
\end{equation}
The determinant $J$ is related to the operator $P$ by:
\begin{equation}
    J(\bar e,0) = \sqrt{\det(\Pd P)}.
\end{equation}

To evaluate $\Pd$, one needs inner products for the different vector spaces involved. The inner products we use are:
\begin{equation}\label{in_prod_sve}
    \begin{split}
        G_s(\de f,\de g) & \equiv |\La|\int dA\de f(\sg) \de g(\sg),  
        \\
        G_v\pr{\de V^{\al},\de W^{\bb}} & \equiv \int dAg_{\al\bb}\de V^{\al}(\sg) \de W^{\bb}(\sg), 
        \\
        G_e\pr{\de e^a_{\al},\de f^b_{\bb}} & \equiv \int dAg^{\al\bb}\de _{ab}\de e^a_{\al}(\sg) \de f^b_{\bb}(\sg),
        \\
        G_\ttt(\de\ttt_i,\de\rho_j) & \equiv \frac{\de^{ij}\de\ttt_i\de\rho_j}{\tw^2},
        \\
        G_{\sg}(\de\sg_1^{\al},\de{\sg'}_1^{\bb}) & \equiv g_{\al\bb}(0)\de\sg_1^{\al}\de{\sg'}_1^{\bb}.
    \end{split}
\end{equation}
$G_s$ is used for $\de\Om,\de\phi$, and can be obtained from $G_e$ by considering how the vielbein is changing due to a change in $\Om,\phi$: $\de e^a_{\al} = e^a_\al\de\Om + \ep^a_b e^b_\al\de\phi$. The same is true for $G_\ttt$. If one writes $e=e_pM$ as in \eqref{eq_vilebein_sg}, then:
\begin{equation}
    \int d^2\sg \sqrt{g}g^{\al\bb}\pd_{\ttt_i}e^a_\al\pd_{\ttt_j}e^a_\bb\de\ttt_i\de\rho_j = A\Tr\pr{\pd_{\ttt_i}M M^{-1}T^{-M}\pd_{\ttt_j}M}\de\ttt_i\de\rho_j = \frac{A\de^{ij}\de\ttt_i\de\rho_j}{\tw^2},
\end{equation}
where $A$ is the WS-area, and can be removed since it is an overall constant. This inner product is also expected since it is modular invariant. It leads to the invariant measure $\frac{d^2\ttt}{\tw^2}$.\footnote{It is different from the simpler inner product used in \cite{Dubovsky:2018bmo}, which is not modular invariant.}

$P^\dagger$ is given by:
\begin{equation}\label{Pd_def}
    \Pd
    \begin{pmatrix} 
        \de e_{a\g} \\ \de\sg_1^\al 
    \end{pmatrix}
    =
    \begin{pmatrix}
        \bar{g}^{\bb\de}\bar{g}^{\al\g}(-\bar{e}^a_{\bb}\pd_{\al}-\pd_{\al}\bar{e}^a_{\bb}+\pd_{\bb}\bar{e}^a_{\al})
        & \frac{\de(\sg)}{\sqrt{\bar{g}(0)}}\de_\al^\de
        \\
        -\bar{g}^{\al\g}\bar{e}^a_{\al} & 0 
        \\
        -\bar{g}^{\al\g}\ep^a_b \bar{e}^b_{\al} & 0
        \\
        -\de_{kj}\tw^2\int d^2\sg\sqrt{g}g^{\g\bb}\pd_{\ttt_j}\bar{e}^a_\bb & 0
    \end{pmatrix}
    \begin{pmatrix} 
        \de e_{a\g} \\ \de\sg_1^\al 
    \end{pmatrix}.
\end{equation}
Composing $P^\dagger P$ gives:
\begin{equation}\label{PdP_def}
\Pd P =
    \begin{pmatrix}
        {P_{11}}^{\de}_{\bb}
        & \pd^{\de}
        & \ee^{\de\al}\pd_{\al} + 2\ee^{\de\al}\pd_{\al}\Om -2\pd^{\de}\phi 
        & {P_{14}}^{\de i}
        \\
        -2\pd_{\bb}\Om-\pd_{\bb}
        & 2 
        & 0  
        & a_i
        \\
        -2\pd_{\bb}\phi-{\ee_{\bb}}^\al\pd_{\al} 
        & 0
        & 2
        & b_i
        \\
        {P_{41}}_{k\bb} 
        & \tw^2 a_k\int d^2\sg\sqrt{g}
        & \tw^2 b_k\int d^2\sg\sqrt{g}
        & \de_k^i\tw^2\bar A
    \end{pmatrix},
\end{equation}
where
\begin{equation}
    \begin{split}
        {P_{11}}^{\de}_{\bb} \equiv & -\de^{\de}_{\bb} g^{\al\g}\pd_{\al}\pd_{\g} -(\de^{\de}_{\al}\pd_{\bb}\Om+2\de^{\de}_{\bb}\pd_{\al}\Om+\ee^{\de}_{\al}\pd_{\bb}\phi-\ee_{\bb\al}\pd^{\de}\phi)\pd^{\al}+\pd^{\de}\Om\pd_{\bb} \\
        & +2\pd^{\de}\phi\pd_{\bb}\phi -2\ee^{\de}_{\al}\pd^{\al}\Om\pd_{\bb}\phi-\pd^{\de}\pd_{\bb}\Om-\ee^{\de}_{\al}\pd^{\al}\pd_{\bb}\phi + \frac{\de(\sg)}{\sqrt{\bar{g}(0)}}\de^{\de}_{\bb},
        \\
        {P_{14}}^{\de i} \equiv & \bar{g}^{\bb\de}\bar{g}^{\al\g}(\bar{e}^a_{\bb}\pd_{\al}\pd_{\ttt_i}\bar{e}^a_\g + (\pd_{\al}\bar{e}^a_{\bb} - \pd_{\bb}\bar{e}^a_{\al})\pd_{\ttt_i}\bar{e}^a_\g),
        \\
        {P_{41}}_{k\bb} \equiv & -\de_{kj}\tw^2\int d^2\sg\sqrt{g}g^{\g\de}\pd_{\ttt_j}\bar{e}^a_\de\pr{\pd_{\bb}\bar{e}^a_{\g} + \bar{e}^a_{\bb}\pd_{\g}},
        \\
        a_i \equiv & \bar{g}^{\al\g}\bar{e}^a_{\al}\pd_{\ttt_i}\bar{e}^a_\g = \Tr\pr{M^{-T}\pd_{\ttt_i}M^T},
        \\
        b_i \equiv & \bar{g}^{\al\g}\ep^a_b \bar{e}^b_{\al}\pd_{\ttt_i}\bar{e}^a_\g = \Tr\pr{\ep M^{-T}\pd_{\ttt_i}M^T}.
    \end{split}
\end{equation}

One expects $J$ to have the following properties:
\begin{enumerate}
    \item $J \propto \det\pr{\frac{e^{-2\Om}}{\tw}}$. This factor will cancel the extra $\det\pr{e^{2f_\Om}\tw}$ in \eqref{eq_dev_cor_1}. The first row of $\Pd P$ is multiplied by $e^{-2\Om}$, and it will be pulled out as $\det\pr{e^{-2\Om}}$ when taking the determinant.\footnote{Let $M$ be a $n \cross n$ block matrix, and say its first row is $m_{1j} = ea_{1j}$, each one being a matrix. Then $M = EM'$, where $e_{11} = e,\ e_{ii} = I,\ e_{ij} = 0$, and $m'_{1j} = a_{1j},\ m'_{ij} = m_{ij}$. Taking determinant of both sides gives $\det(M) = \det(E)\det(M') = \det(e)\det(M')$. The differential diffeomorphism space is two dimensional, so $\det(e) = \det\pr{e^{-2\Om}}^2$, and the square root leaves $\det\pr{e^{-2\Om}}$. We expect $e^{-2\Om}$ to be accompanied by $\tw^{-1}$, as they appear together in $\sqrt{g}$.}

    \item $J \propto \frac{1}{\tw^2}$. The correlation function should be modular invariant by changing $\ttt$ and $T$ simultaneously. The FP-identity \eqref{eq_FP_identity} comes without $T$ and, therefore, should be modular invariant with respect to $\ttt$ alone. Given the integration measure $d^2\ttt$, the combination $\frac{d^2\ttt}{\tw^2}$ is modular invariant (it indeed happened for the partition function \cite{Dubovsky:2018bmo}).\footnote{It does not mean that it is the only dependence of $\tw$ inside $J$, the WS-area $\bar A = \tw\int d^2\sg e^{\Om}$ is also modular invariant. It just means that only in this factor $\tw$ will appear alone.}

    \item $J$ is independent of the momentum at large momentum. This is because $P^\dagger P$, after taking $e^{-2\Om}$ from the first row, depends on $\Om$ (and $\phi$) through derivatives only. As large momentum, using \eqref{eq_z_sol} and the finite values for the saddle of $\bar\Om,\bar\phi$ given in subsection \ref{subsec_sad_constnat}:
    \begin{equation}
        e^{\Om - i\phi} \approx \sum_i Q_i\pd G_i \equiv K\sum_i \tilde Q_i\pd G_i,
    \end{equation}
    where we pulled out the large momentum scale $K$. Then:
    \begin{equation}
    \begin{split}
        \Om &\approx \ln(K) + \ln\pr{\left| \sum_i \tilde Q_i\pd G_i \right|}, 
        \\
        \phi &\approx -\frac{i}{2}\ln\pr{\frac{\sum_i \bar{\tilde{Q}}_i\bpd G_i}{\sum_i \tilde{Q}_i\pd G_i}},
    \end{split}
    \end{equation}
    and the dependence on $K$ disappear when taking the derivatives.

\end{enumerate}


\subsection{Q}

The operator $Q$ is the linearization of the constraint $K_a(\sg) = 0$ around its solution $f_\Om,f_\phi$. It is given by:
\begin{equation}
    Q
    \begin{pmatrix}
        \de\Om \\ \de\phi    
    \end{pmatrix}
    \equiv \ep^{\al\bb}\left(\pd_\al\bar{e}^c_{\bb} + \bar{e}^c_{\bb}\pd_\al \right)
    \begin{pmatrix}
        \de^b_c & \ep^b_c
    \end{pmatrix}
    \begin{pmatrix}
        \de\Om \\ \de\phi    
    \end{pmatrix}.
\end{equation}
It has two zero modes, as the general solution has two free parameters. It is easy to verify that $Q$ indeed annihilates the two vectors:
\begin{equation}\label{eigenvectors_Q}
    e^{\bar\Om - f_\Om}
    \begin{pmatrix}
        \cos\left(f_\phi - \bar\phi\right) \\
        -\sin\left(f_\phi - \bar\phi\right)
    \end{pmatrix}
    ,\ 
    e^{\bar\Om - f_\Om}
    \begin{pmatrix}
        \sin\left(f_\phi - \bar\phi\right) \\
        \cos\left(f_\phi - \bar\phi\right)
    \end{pmatrix}.
\end{equation}

We use the inner product out of which the discrete measure $\prod_{\sg=1}^N d\Om(\sg) d\phi(\sg)$ is naturally defined (see subsection \ref{subsec_conf-ori-PI}), which is $\sum_{\sg=1}^N\pr{d\Om(\sg)d\Om'(\sg) + d\phi(\sg)d\phi'(\sg)}$. Then $Q^{\dagger}$ is:
\begin{equation}
    Q^{\dagger}\de K_b \equiv 
    -
    \begin{pmatrix}
        \de^b_c \\
        \ep^b_c
    \end{pmatrix}    \ep^{\al\bb}\bar{e}^c_{\bb}\pd_{\al}\de K_b.
\end{equation}
The composition $Q^{\dagger}Q$ gives:
\begin{equation}
\begin{split}
    Q^{\dagger}Q &= 
    \begin{pmatrix}
        \ A & B\ \\
        -B & A\ 
    \end{pmatrix}, 
    \\
    A &\equiv -g g^{\al\bb} \pr{ \pd_\al\pd_\bb + \pd_\al\Om\pd_\bb + \pd_\bb\Om\pd_\al + \pd_\al\Om\pd_\bb\Om - \pd_\al\phi\pd_\bb\phi +\pd_\al\pd_\bb\Om}, 
    \\
    B &\equiv g g^{\al\bb} \left(\pd_\al\phi\pd_\bb + \pd_\bb\phi\pd_\al + \pd_\al\phi\pd_\bb\Om + \pd_\bb\phi\pd_\al\Om + \pd_\al\pd_\bb\phi \right),
\end{split}
\end{equation}
where $g = \det(g_{\al\bb}) = e^{4\Om}\tw^2$.\footnote{In \cite{Aharony:2023dod} another term appeared in $A$: $2g\ee^{\al\bb}\pd_\al\Om\pd_\bb\phi,$. It was a mistake, reminiscent of old versions.} By deriving this result, the following identities were used:
\begin{equation}    
    \ep^{\g\de}\ep^{\al\bb}e^b_\bb e^c_\de \begin{pmatrix}
        \de_{cb} \\ \ep_{cb}
    \end{pmatrix} 
    = g
    \begin{pmatrix}
        g^{\g\al} \\ \ee^{\g\al} 
    \end{pmatrix},
\end{equation}
which are valid in general, particularly both for the plane and for the torus. The diagonal form of $Q^{\dagger}Q$ is:
\begin{equation}
\begin{split}
    \begin{pmatrix}
        \ A & B\ \\
        -B & A\ 
    \end{pmatrix} 
    & =
    U
    \begin{pmatrix}
        A+iB & 0 \\
        0 & A-iB \ 
    \end{pmatrix}
    U^\dagger ,
    \\
    U & \equiv \frac{1}{\sqrt{2}}
    \begin{pmatrix}
        I & I\ \\
        iI & -iI\ 
    \end{pmatrix}.
\end{split}
\end{equation}
The explicit form of the operators $A\pm iB$ can be simplified using $\bar z = e^{\Om-i\phi}$:
\begin{equation}\label{eq_det_Q_gen}
    \sqrt{{\det}'(Q^{\dagger}Q)} = \left|{\det}'\pr{-gg^{\al\bb}(\pd_\al +\pd_\al\ln(\bar z))(\bpd_\bb + \bpd_\bb\ln(\bar z))}\right|.
\end{equation}
The form of the operator inside the determinant looks nicer when using complex coordinates:
\begin{equation}\label{eq_det_Q_w}
    gg^{\al\bb}(\pd_\al +\pd_\al\ln(\bar z))(\bpd_\bb + \bpd_\bb\ln(\bar z)) = e^{2\Om}(\pd +\pd\ln(\bar z))(\bpd + \bpd\ln(\bar z)).
\end{equation}
However, the form of the constraints $K_b(\sg)$ was obtained using the $\sg$ coordinates, and the complex coordinates $w$ depend explicitly on $\ttt$, which is integrated, so this determinant should be thought of in terms of $\sg$ coordinates only.

\eqref{eq_det_Q_w} suggest to take $\det(e^{2\Om }\tw)$ outside of \eqref{eq_det_Q_gen}. But, because of the zero modes exclusion, we can not naively split ${\det}'\left(e^{2\Om}\tw\square \right) \rightarrow {\det} \left(e^{2\Om}\tw\right) {\det}'(\square)$, where $\square \equiv \frac{1}{\tw}(\pd +\pd\ln(\bar z))(\bpd + \bpd\ln(\bar z))$. In order to ``resolve" the zero modes, the operator $D\equiv e^{2\Om}\tw\square$ is regulated: $D_\ep \equiv e^{2\Om}\tw(\square + \ep)$, where $\ep$ is a small number. The eigenvalues $\lambda^\ep_n$ of $D_\ep$ are slightly shifted from $\lambda_n$ of $D$: $\lambda^\ep_n \approx \lambda_n + \ep \al_n$. The eigenvalue $\lambda_0 = 0$ was excluded in ${\det}'(D)$, and therefore $\lambda^\ep_0 \approx 0 + \ep\al_0$ will be excluded in the regularized version. Using these definitions, we evaluate:
\begin{equation}
\begin{split}
    {\det}'(D_\ep) &= \prod_{n\neq 0}\lambda^\ep_n = \frac{\prod_{n}\lambda^\ep_n}{\lambda^\ep_0} = \frac{\det(D_\ep)}{\ep\al_0} = \frac{\det(e^{2\Om}\tw)\det(\square + \ep)}{\ep\al_0} = \det(e^{2\Om}\tw) \frac{{\det}'(\square + \ep)\ep}{\ep\al_0}  \\
    &\xrightarrow[{\ep \rightarrow 0}]{} {\det}'(D) =  \frac{1}{\al_0}\det(e^{2\Om}\tw){\det}'(\square).
\end{split}
\end{equation}
The finite factor $\al_0^{-1}$ can be calculated as follows. We write the eigenfunction $f^\ep_0 \approx \frac{1}{\bar z} + \ep f_0$, and solve the eigenvalue equation:
\begin{equation}\label{det_Q_2}
e^{2\Om}\left(\square + \ep\right)f^\ep_0 = \lambda^\ep_0 f^\ep_0 \Rightarrow \pd\bpd(\bar z f_0) = \frac{\al_0}{e^{2\Om}\tw} - 1.
\end{equation}
$f_0$ and $\bar z$ satisfy periodic boundary conditions, therefore:
\begin{equation}
    \int d^2\sg \pd\bpd(\bar z f_0) = 0 \Rightarrow \al_0 = \frac{\tw}{\int d^2\sg e^{-2\Om}}.
\end{equation}
Finally, the determinant can be written as:
\begin{equation}\label{det_dec_3} 
    \sqrt{{\det}'(Q^{\dagger}Q)} 
    = \det(e^{2\Om}\tw) \frac{\int d^2\sg e^{-2\Om}}{\tw} \left|{\det}'\pr{-\frac{1}{\tw}(\pd +\pd\ln(\bar z))(\bpd + \bpd\ln(\bar z))}\right|.
\end{equation}


\newpage
\section{Special Functions}\label{app_fun}
\subsection{Theta-function}

The $\Th(w,\ttt)$-function is defined as:
\begin{equation}
\begin{split}
    \Th(w,\ttt) &\equiv i \sum_{n= -\infty}^\infty (-1)^n q^{\frac{(n-\frac{1}{2})^2}{2}} z^{n-\frac{1}{2}}
    \\
    &= 2 q^{\frac{1}{8}}\sin(\pi w) \prod_{m=1}^\infty (1-q^m)(1-zq^m)(1-z^{-1}q^m),
    \\
\end{split}
\end{equation}
where:
\begin{equation}
    q \equiv e^{2\pi i\ttt},\ z \equiv e^{2\pi iw}.
\end{equation}
$\Th(w,\ttt)$ here is $EllipticThetaPrime[1,z',q']$ in Mathematica, with $q' = \sqrt{q} = e^{\pi i\ttt}$ and $z' = \frac{-i\ln(z)}{2} = \pi w$. It was used in the numerical calculation done in section \ref{sec_sad}. In the literature it is called $-\Th_{11}(w,\ttt) = -\Th\genfrac[]{0pt}{2}{\frac{1}{2}}{\frac{1}{2}}(w,\ttt)$. There are various conventions for the variables and definitions of $z,q$, so one has to be careful when comparing expressions from different sources. We list its important properties:
\begin{align}
    \Th(0,\ttt) &= 0,
    \\
    \Th(-w,\ttt) &= -\Th(w,\ttt),
    \\
    \overline{\Th(w,\ttt)} &= -\Th(-\bar w,-\bar\ttt) = \Th(\bar w,-\bar\ttt),
    \\
    \Th(w,\ttt) &\approx 2\eta(\ttt)^3\pi w\ |w|<<1.\label{eq_Theta_small_w}
\end{align}
It is not periodic:
\begin{align}
    \Th(w+1,\ttt) &= -\Th(w,\ttt), 
    \\
    \Th(w+\ttt,\ttt) &= -e^{-\pi i\ttt-2\pi iw}\Th(w,\ttt).
\end{align}    
Its modular properties are:
\begin{equation}
\begin{split}
    \Th(w,\ttt+1) &= e^{\frac{i\pi}{4}} \Th(w,\ttt),
    \\
    \Th\pr{\frac{w}{\ttt},-\frac{1}{\ttt}} &= -i\sqrt{-i\ttt} e^{\frac{i\pi w^2}{\ttt}} \Th(w,\ttt).
\end{split}
\end{equation}

\subsection{Eta-function}

The $\eta(\ttt)$-function is defined as:
\begin{equation}
    \eta(\ttt) \equiv q^{\frac{1}{24}} \prod_{m=1}^\infty (1-q^m),
\end{equation}
with $q$ as before. $\eta(\ttt)$ here is $DedekindEta[\ttt]$ in Mathematica. It was used in the numerical calculation done section \ref{sec_sad}. Its modular properties are:
\begin{equation}
\begin{split}
    \eta(\ttt+1) &= e^{\frac{i\pi}{12}} \eta(\ttt),
    \\
    \eta\pr{-\frac{1}{\ttt}} &= (-i\ttt)^{\frac{1}{2}} \eta(\ttt).
\end{split}
\end{equation}

\subsection{G-function}

$G$ is naturally defined as a function of $w$ and $\ttt$ as in \eqref{eq_def_G}:
\begin{equation}\label{eq_def_G_app}
    G(w,\ttt) \equiv -\ln\left| \Th\pr{w|\ttt} \right|^2 + \frac{2\pi\Im(w)^2}{\tw}.
\end{equation}
One can replace $w$ by the Cartesian coordinates $\sg^{1,2}$:
\begin{equation}
    G(\sg^1,\sg^2,\ttt) \equiv G(\sg^1 + \ttt\sg^2,\ttt).
\end{equation}
It satisfies:
\begin{align}
    G(-w,\ttt) &= G(w,\ttt),
    \\
    \overline{G(w,\ttt)} &= G(w,\ttt).
\end{align}
The second term in \eqref{eq_def_G_app} ``fixes" the non-periodicity of the $\Th$-function:
\begin{align}
    G(w+1,\ttt) = G(w,\ttt) &\Rightarrow G(\sg^1 + 1,\sg^2,\ttt) = G(\sg^1,\sg^2,\ttt),
    \\
    G(w + \ttt,\ttt) = G(w,\ttt) &\Rightarrow G(\sg^1,\sg^2 + 1,\ttt) = G(\sg^1,\sg^2,\ttt).
\end{align}
The modular properties of $G$ are dictated by those of $\Th$:
\begin{equation}
\begin{split}
    G(w,\ttt) = G(w,\ttt+1) &\Rightarrow G(\sg^1,\sg^2,\ttt) = G(\sg^1-\sg^2,\sg^2,\ttt+1),
    \\
    G\pr{\frac{w}{\ttt},-\frac{1}{\ttt}} = -\ln{|\ttt|} + G(w,\ttt) &\Rightarrow G\pr{\sg^2,-\sg^1,-\frac{1}{\ttt}} = -\ln{|\ttt|} + G(\sg^1,\sg^2,\ttt).
\end{split}
\end{equation}
When changing the second argument of $G(w,\ttt)$ to $f_3(\ttt)$, the first argument should be in the form $f_1(\sg^{1,2}) + f_2(\sg^{1,2})f_3(\ttt)$, and then identify the dependence in Cartesian coordinates $G(f_1(\sg^{1,2}),f_2(\sg^{1,2}),f_3(\ttt))$.

The divergences of $G$ come solely from zeros of $\Th$. The latter has a unique zero at $\sg^1 = \sg^2 = 0$, so this is the only divergence of $G$. The value $\tw = 0$ is also problematic, but it is not allowed.

At $\tw \gtrsim 2$, $G$ grow linearly with $\tw$. Three terms contribute to the slope: $q^\frac{1}{8}$ and $\sinh(\pi w)$ from the first term in the r.h.s. of \eqref{eq_def_G_app}, and $2\pi\tw(\sg^2)^2$ from the second term:
\begin{equation}\label{eq_large_tw_G}
    G(\sg^1,\sg^2,\ttt) \approx 2\pi\pr{\sg^2-\frac{1}{2}}^2\tw.
\end{equation}
The small $\tw$ behavior is more subtle and can be derived by the modular properties of $G$. 


\newpage
\section{Weyl Anomaly}\label{app_weyl}
In subsection \ref{subsec_Undeformed-PI}, the Weyl anomaly for a conformal change $g = e^{2\Om}g_0$ of a given initial metric $g_0$ was needed. It can be calculated in similar steps to those presented in \cite{Polchinski:1998rq}. The following differential equation for a partition function $Z[g]$ of a QFT defined on a curved space with the metric $g$ holds:
\begin{equation}\label{eq_weyl_conf_1}
    \frac{\de Z\left[g\right]}{\de g_{\al\bb}(\sg)} = -\frac{1}{2}\sqrt{g(\sg)}Z[g]\langle T^{\al\bb}(\sg)\rangle.
\end{equation}
Using $\de g_{\al\bb} =  2g_{\al\bb}\de\Om$ gives:
\begin{equation}\label{eq_weyl_conf_2}
    \frac{\de Z\left[e^{2\Om}g_0\right]}{\de \Om(\sg)} = -e^{2\Om}\sqrt{g_0(\sg)}Z\left[e^{2\Om}g_0\right]\langle T^\al_\al(\sg)\rangle.
\end{equation}
For a manifold without a boundary, the trace of the energy-momentum tensor is $-\frac{c}{12}R$. If $g_0$ is a flat metric, the Ricci scalar for $g$ is:
\begin{equation}
    R = -2e^{-2\Om}\nabla^2_{g_0}\Om.
\end{equation}
Inserting these two relations into \eqref{eq_weyl_conf_2} yields:
\begin{equation}
    \frac{\de Z\left[e^{2\Om}g_0\right]}{\de \Om(\sg)} = -\frac{c}{6}\sqrt{g_0(\sg)}Z\left[e^{2\Om}g_0\right]\nabla^2_{g_0}\Om.
\end{equation}
This equation can be integrated:
\begin{equation}\label{eq_weyl_conf_3}
    Z\left[e^{2\Om}g_0\right] = Z\left[g_0\right]e^{-\frac{c}{12}\int d^2\sg\sqrt{g_0(\sg)}\Om\nabla^2_{g_0}\Om} \equiv Z\left[g_0\right]e^W.
\end{equation}
Integration by parts gives \eqref{eq_weyl_1}:
\begin{equation}
    W = \frac{c}{12}\int d^2\sg \sqrt{g_0}g_0^{\al\bb}\pd_\al\Om\pd_\bb\Om.
\end{equation}
In complex coordinates, it reads:
\begin{equation}\label{eq_weyl_conf_4}
    W = \frac{c}{6}\int d^2w\sqrt{g_0(\sg)}\pd\Om\bpd\Om.
\end{equation}
The measure $d^2w$ is nothing than the Cartesian measure, it is convenient to use it when integrating $\de(w)$.

We evaluate it for $\Om = \frac{1}{2}(\ln(z) + \ln(\bar z))$. The symmetric form $\pd\Om\bpd\Om$ is preferred from $\Om\pd\bpd\Om$, since these two derivatives is ambiguous for the $\ln(w),\ln(\bar w)$ inside $\Om$. 

\eqref{eq_weyl_conf_4} yields three terms, all multiplied by $\frac{c}{24}$. 
\newline The first is:
\begin{equation}
\begin{split}
    \int d^2w \frac{\bpd \bar z \pd \bar z}{\bar z^2} + c.c. & = -2\pi\sum_{i=1}^2 Q^i\frac{\pd \bar z(w_i)}{\bar z^2(w_i)} + c.c. 
    \\
    & = -\sum_{i=1}^2 Q^i \int_0^{2\pi} d\al \frac{Q^i/(\ep e^{i\al})^2}{(Q^i)^2/(\ep e^{i\al})^2} + c.c. = -\sum_{i=1}^2 4\pi.
\end{split}
\end{equation}
\newline The second is:
\begin{equation}
    \int d^2w \frac{\pd z \bpd \bar z}{|z|^2} = \int d^2w \frac{4\pi^2}{|z|^2}\sum_{i,j=1} Q^i \bar Q^j\de(w-w_i)\de(w-w_j) = 0.
\end{equation}
For $i \neq j$, it vanishes immediately, but it also vanishes for $i=j$ because of the point-splitting regularization.
\newline The third is:
\begin{equation}\label{Weyl_3}
    \int d^2w \frac{\bpd z \pd \bar z}{|z|^2} = \int d^2w \frac{1}{|z|^2} \sum_{i,j=1}^2 Q^i \bar Q^j \pd^2G_i(w)\bpd^2G_j(w).
\end{equation}
The factor $\sum_{i,j=1}^2 Q^i \bar Q^j \pd^2G_i(w)\bpd^2G_j(w)$, gives poles at $w_i$ for all $i$. As $w$ approaches $w_i$, $\pd^2G_i(w) \sim \frac{1}{(w-w_i)^2}$. Hence the integrand becomes $|w-w_i|^{-2}$, which gives a logarithmic divergence $-2\pi\ln(\ep)$ (the integral of terms with $j\neq i$ near $w_i$ vanishes due to the point-splitting). The factor $\frac{1}{|z|^2}$ gives poles at the zeros of $\bar z$, which we denote as $p_i$. The residue at these poles is $1$.\footnote{The sum of all residues of $\frac{\pd \bar z}{\bar z}$ vanishes, consistent with the fact that it is periodic.} Each of these poles will contribute another logarithmic divergence $-2\pi\ln(\ep)$. Therefore, using some arbitrary cut $\ep<|w-w_i|,|w-p_i|<r$ around each pole, the integral of $\left|\frac{\pd\bar z}{\bar z}\right|^2$ is $2\sum_{i=1}^2 4\pi\ln \left( \frac{r}{\ep} \right) + I_0$, where $I_0$ is finite and depends on the momenta $Q^i$ and the positions $w_i$. The divergent part can be deduced in the following way. Since the point splitting is imposed for $|w-w_i|,|w-p_i|$ around each pole, it is convenient to change variables:
\begin{align}
    {\sg^1}'=\sg^1+\tn \sg^2,\ {\sg^2}'=\tw \sg^2,\ 
    d{\sg^1}'d{\sg^2}'=\tw d\sg^1 d\sg^2,
    \\
    r' = \sqrt{\pr{{\sg^1}'}^2+\pr{{\sg^2}'}^2},\ \phi'=\tan^{-1}\pr{\frac{{\sg^2}'}{{\sg^1}'}}, \ r'dr'd\phi'=d{\sg^1}'d{\sg^2}',
\end{align}
and then the integral reads:
\begin{equation}
    \int_{\ep<|w|<r} \frac{d^2w}{|w|^2} = \int_{\ep<r'<r} \frac{2\tw d\sg^1 d\sg^2}{r'^2} = 2\int_{\ep<r'<r} \frac{r'dr'd\phi'}{r'^2} = 4\pi\ln\pr{\frac{r}{\ep}}.
\end{equation}
In conclusion, $W$ takes the form:
\begin{equation}\label{Weyl_5}
    W = \frac{c}{24} \left( \sum_{i=1}^2 8\pi\ln \left( \frac{r}{\ep} \right) + I_0 - 4\pi\right).
\end{equation}
Upon exponentiation $e^W$ gives:
\begin{equation}\label{Weyl_6}
    e^{\frac{cI_0}{24} - \frac{\pi c}{6}}\prod_{i=1}^2 \left(\frac{r}{\ep} \right)^\frac{\pi c}{3}.
\end{equation}
A divergent part $\ep^{-\frac{\pi c}{3}}$ is present for each operator. $I_0$ is independent of the momentum at large momentum and, therefore, is not relevant for the saddle approximation.\footnote{The argument given in \cite{Aharony:2023dod} pp.25 why $I_0$ is not relevant for the saddle is wrong, since $I_0$ is inside an exponent. The right argument is given here.}


\newpage
\section{Saddle on The Complex Plane}\label{app_com}
In this appendix, we briefly explain how the saddle point approximation is used for complex integration. We illustrate it for a simple example of a Gaussian integral. Then we show how it works for the high momentum limit of the plane and the torus.

\subsection{contour integration}

The integral $\int_{-\infty}^\infty dx e^{Nf(x)}$ can be evaluated by analytic continuing $f(x)\to f(w),\ w \equiv x+iy$ to the complex plane. Then, assuming $e^{Nf}$ decays fast enough, one can turn the line into a closed contour and deform it as wanted. If it crosses any poles / branch-cuts, then their contribution should be collected.

For using the saddle point approximation at large $N$, one wants to avoid integrating over a changing phase. It is preferred that the phase of the integrand will be constant along the integration contour. The way to find a point $w_0$ suitable for a saddle is by solving $\frac{df(w)}{dw}=0$. Since $f(w)$ is an analytic function, $\frac{df(w)}{dw}=0$ guarantees that as two functions of two variables $\Re f(x,y),\Im f(x,y)$, all of the directional derivatives vanish at $w_0$. In particular, this is a stationary point of the phase and the modulus of $e^{Nf}$.

For any analytic function with the expansion $g(w_0+dw) \approx g(w_0) + \frac{g^{(n)}(w_0)}{n!}\cdot dw^n$ where the first $n-1$ derivatives vanish at $w_0$, one can scan the direction $\phi$ around $w_0$ with $dw = \ep e^{i\phi}$ for small $\ep$. Then:
\begin{equation}
    g(w_0,\phi) \approx g(w_0) + \frac{g^{(n)}(w_0)}{n!}\ep^n e^{in\phi}.
\end{equation}
The phase of $g$ is:
\begin{equation}
    \arg(g(w_0+dw)) \approx \arg(g(w_0)) + \ep^n \pr{c_1\cos{n\phi} + c_2\sin{n\phi}},
\end{equation}
where $c_{a,2}$ are constants. Hence, the phase will change a bit, and as $dw$ makes a full round $g(w_0+dw)$ makes $n$-rounds. In particular, for $n=1$, which is a regular point, each allowed phase (which is $\ep$ close to $\arg(g(w_0)$) will appear twice (except the two extreme phases that will appear once). For $n=2$, when the first derivative vanishes, each phase will appear four times (except the two extreme phases, which will appear twice). Therefore, around the saddle, which in general has $n=2$, there will be two lines (four directions) going out $g(w_0)$ with the same phase. The modulus will also be extremized at $w_0$ along each line, and we want a line along which it is a maximum. 

\subsection{Gaussian Integral}

As a simple example, we consider the Gaussian integral $\int dq e^{iqx-q^2}$
at large $x$. We expect from naive considerations to find saddle at $q_0 \sim x$. Indeed:
\begin{equation}
    \frac{d}{dq}\pr{iqx-q^2} = 0 \Rightarrow q_0 = \frac{ix}{2}.
\end{equation}
Plugging in the integral gives $e^{-\frac{x^2}{4}}$, which is the saddle approximation of this integral. To show this through the line of reasoning presented above, we find the lines of constant phase emerging from the saddle. The imaginary part at $\frac{ix}{2}$ is zero. Setting $q=k+il$, one solves the following equation:
\begin{equation}
    \Im{i(k+il)x - (k+il)^2)} = 0.
\end{equation}
There are two solutions. The first is $k=0$, and the second is $l=\frac{x}{2}$. The two contours of constant phase are $q=il$ and $q=k+i\frac{x}{2}$. Along the first contour, the function is minimized at $\frac{x}{2}$, while along the second contour, it is maximized. Therefore we choose the second contour. The full contour can be completed with two segments of finite imaginary part and real part going to infinity\footnote{The contour can not be completed as usual with great upper / lower half circles.}:
\begin{equation}
\begin{split}
    C_1 &= \{k,\ -r \leq k \leq r \}, 
    \\
    C_2 &= \{r+ik,\ 0 \leq k \leq \frac{x}{2} \}, 
    \\
    C_3 &= \{(-k+i\frac{x}{2}),\ -r \leq k \leq r \},
    \\
    C_4 &= \{-r-i\pr{\frac{x}{2}-k},\ 0 \leq k \leq \frac{x}{2} \}.
\end{split}
\end{equation}
As $r\to\infty$ the contribution of $C_{2,4}$ vanishes. Since the function does not have any poles inside the contour, the sum of $C_{1,3}$ also vanishes. They go in opposite directions, so:
\begin{equation}
    \int_{-\infty}^\infty dq e^{iqx-q^2} = \int_{-\infty}^\infty dk e^{i(k+ix/2)x-(k+ix/2)^2} =   e^{i0} \int_{-\infty}^\infty dk e^{-\frac{x^2}{4}-k^2}.
\end{equation}
The last integral, now over a real function, can be evaluated using the saddle approximation (in this case, it is trivial, but in general, it can not be solved exactly).

\subsection{Plane}

The high momentum limit of the two-pt function defined on the plane takes the form (see (109) in \cite{Aharony:2023dod}):
\begin{equation}
    I = \int d^2\sg e^{iq\sg - \frac{q^2}{\pi|\La|}\ln\pr{|\sg|}}.
\end{equation}
Without loss of generality, we choose $q>0$ to be in the $\sg^1$-axis. Therefore, the saddle should be with $\sg^2=0$. For $\frac{q^2}{|\La|}>>1$, the integral over $\sg^2$ can be evaluated using saddle point approximation (without any modification as the integrand is real)\footnote{In this case, it can be evaluated exactly by: 
\begin{equation}
    \int d\sg^2 \frac{1}{\pr{\pr{\sg^1}^2 + \pr{\sg^2}^2}^a} = \frac{\sqrt{\pi}\Gamma\pr{a-\frac{1}{2}}}{\Gamma(a)} \frac{|\sg^1|}{{\sg^1}^{2(1-a)}}.
\end{equation}
}
:
\begin{equation}
    I = \int d\sg^2 e^{-\frac{q^2}{2\pi|\La|}\ln\pr{\pr{\sg^1}^2 + \pr{\sg^2}^2}} \approx \sqrt{\frac{2\pi^2|\La|\pr{\sg^1}^2}{q^2}}e^{-\frac{q^2}{2\pi|\La|}\ln\pr{\pr{\sg^1}^2}}.
\end{equation}
It remains to perform the integral over $\sg^1$ is:
\begin{equation}
    I = \int d\sg^1 \sqrt{\frac{2\pi^2|\La|}{q^2}}e^{iq\sg^1 + \pr{-\frac{q^2}{2\pi|\La|}+\frac{1}{2}}\ln\pr{\pr{\sg^1}^2}}.
\end{equation}
We analytic continue $\sg^1 \to w \equiv x+iy$. The singularity at $\sg^1=0$ will be skipped by a small half circle going below it: $C_\ep = \{\ep e^{i\phi},\  \pi\leq\phi\leq 2\pi\}$; this is an implementation of the point-splitting regularization. The part on the real axis is $C_L = \{x,\  \ep\leq|x|\leq R\}$. Another large half circle, on which the integrand decays, makes the integration contour closed: $C_R = \{R e^{i\phi},\  0\leq\phi\leq \pi,\ R\to \infty\}$. The branch-cut of the logarithm is chosen, for reasons that will be clear during the calculation, to be along the positive imaginary axis. Altogether, the integral can be evaluated as:
\begin{equation}
    I = \sqrt{\frac{2\pi^2|\La|}{q^2}}\displaystyle\lim_{R\to\infty} \int_{C_L + C_\ep + C_R} dw e^{iqw + \pr{-\frac{q^2}{2\pi|\La|}+\frac{1}{2}}\ln\pr{w^2}}.
\end{equation}
Once we have a closed contour, we can change its shape, and as long as the contour does not cross the pole at $w=0$ or the branch-cut, the integral remains the same.

\begin{figure}
\captionsetup{singlelinecheck = false, justification=justified}
\includegraphics[scale=0.4]{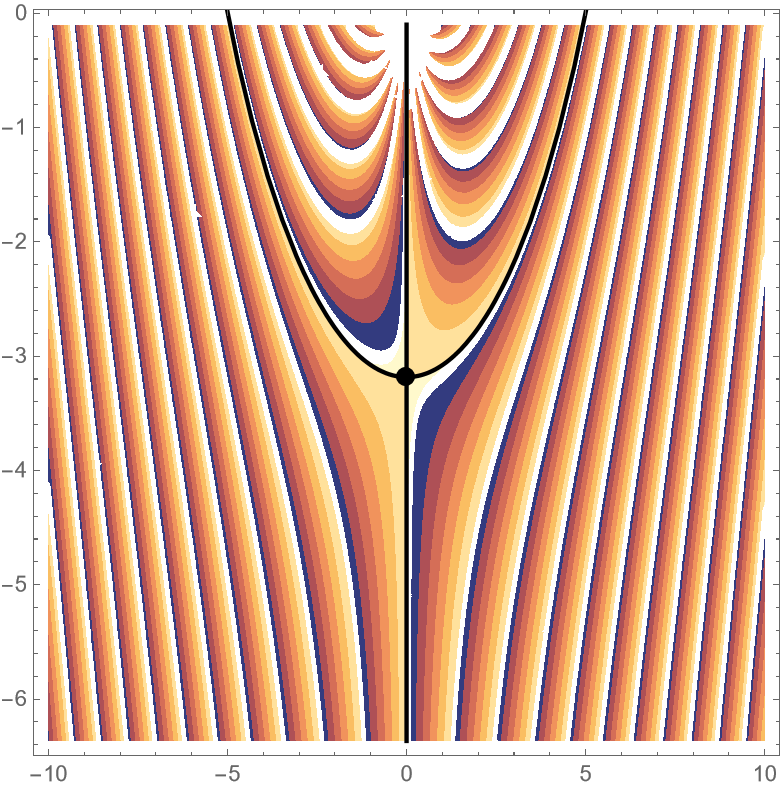}
\quad
\includegraphics[scale=0.4]{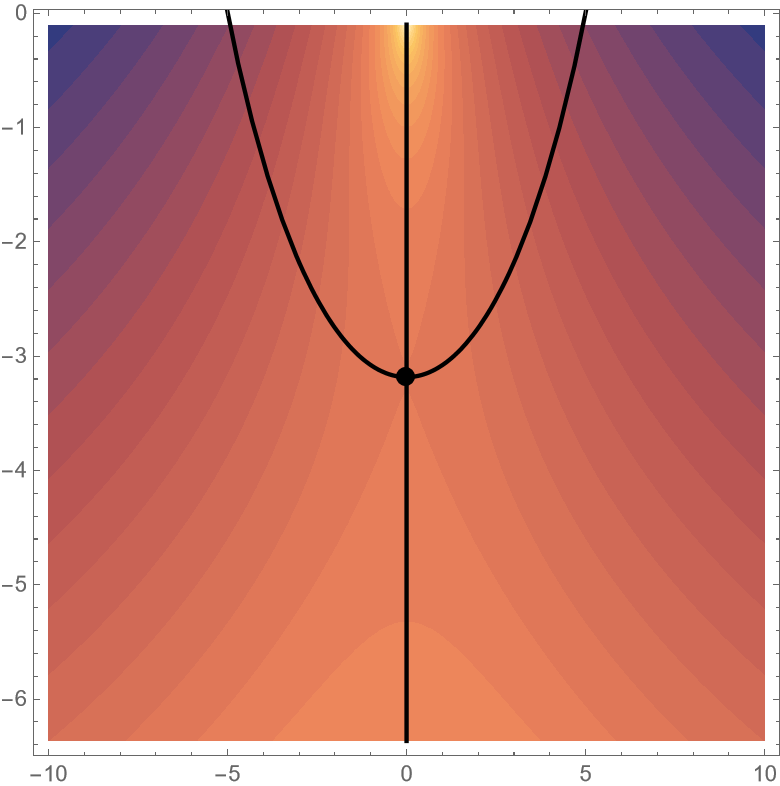}
\caption{The real (left) and imaginary (right) parts of $f(w)$, moded by $2\pi$ ($0$ is identified with $2\pi$). The saddle is shown as a black point. The two contours of constant imaginary part, equals to the imaginary part at the saddle and emerge from it, are shown in black lines. Along one line the saddle is a minimum of the real part, along the other it is a maximum. The latter is the deformed contour.}
\label{fig_im_re_plane}
\end{figure}

\begin{figure}
\captionsetup{singlelinecheck = false, justification=justified}
\includegraphics[scale=0.6]{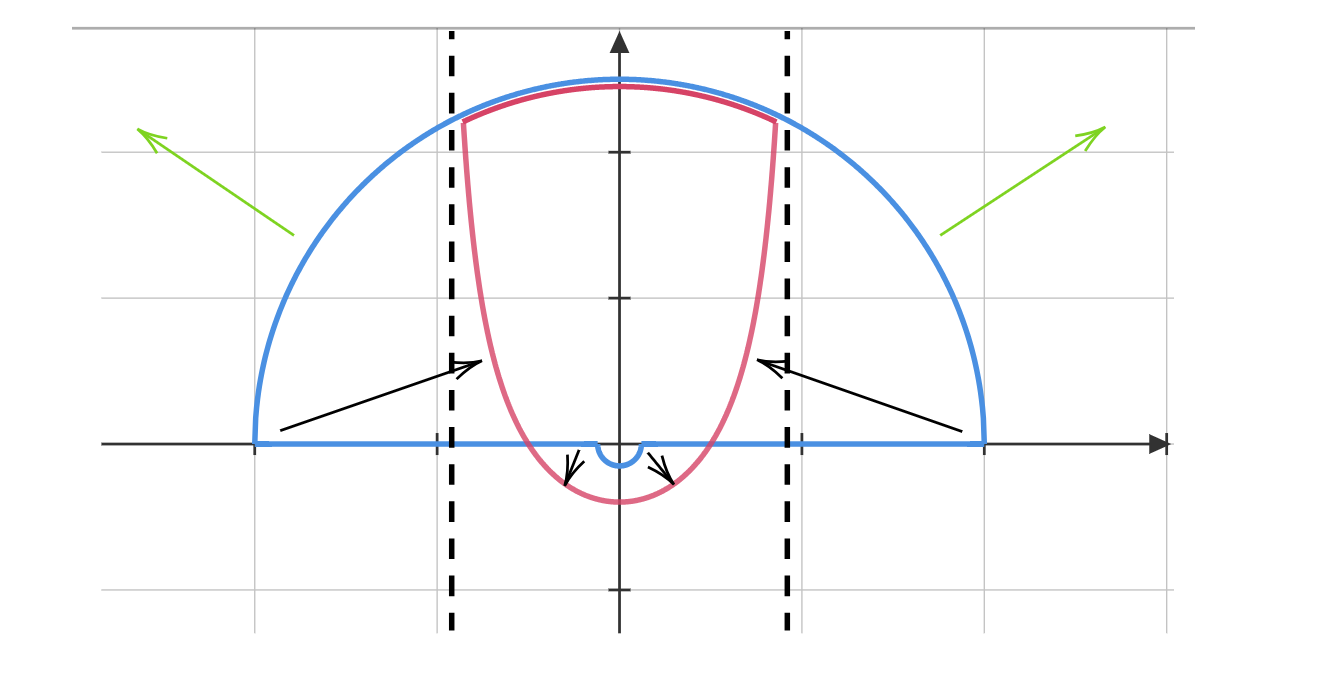}
\caption{Deformation of the integration contour for the plane's correlator. The blue line represents the initial contour, and the red line the deformed contour. The black arrows indicate how the contour is deformed. The green arrows indicate that the radius of the blue contour, as well as the upper part of the red contour, should be taken to infinity. On the deformed contour, the integrand has a constant phase, and its absolute value reaches a maximum at its bottom.}
\label{fig_contours_plane}
\end{figure}

The saddle of $f(w) = iqw -
\frac{q^2}{2\pi|\La|}\ln\pr{w^2}$ is at $w_0 = -i\frac{q}{\pi|\La|}$\footnote{At $\frac{q^2}{|\La|}>>1$, one can use $\frac{q^2}{2\pi|\La|}+\frac{1}{2} \approx \frac{q^2}{2\pi|\La|}$. The extra $\frac{1}{2}$ will give a subleading contribution when substituting the saddle in the integrand.}. The phase of the integrand at the saddle is $e^{-i\frac{q^2}{2|\La|}}$. Two lines of this constant phase pass through the saddle, obeying the equation:
\begin{equation}
    qx - \frac{q^2}{\pi|\La|}\tan^{-1}\pr{\frac{y}{x}} = -\frac{q^2}{2|\La|}.
\end{equation}
The first solution is the imaginary axis: $x=0 \Rightarrow w=iy,\ y<0$. Along this line, however, the function:
\begin{equation}
    \Re f(0+iy) = -qy -\frac{q^2}{\pi|\La|}\ln|y|    
\end{equation}
has a minimum at $y=-\frac{q}{\pi|\La|}$ and not a maximum. The second solution is given by $y(x) = x\tan\pr{\frac{\pi}{2} + \frac{\pi|\La|x}{q}}$. Along this line:
\begin{equation}
    \Re f(x+iy(x)) = q x \cot \left(\frac{\pi|\Lambda|x}{q}\right) + \frac{q^2}{\pi|\Lambda|}\ln \left|\frac{\sin \left(\frac{\pi|\Lambda|x}{q}\right)}{x}\right|.
\end{equation}
It is manifestly symmetric with respect to reflection of $x$ and gets a maximum at $x=0$. Starting from $x=0, y=-\frac{q}{\pi|\La|}$, $y$ grows to infinity as $x$ arrives to $\pm \frac{q}{|\La|}$. $\Re f$ and $\Im f$ are shown in figure \ref{fig_im_re_plane}. These two contours of constant imaginary part emerging from the saddle are shown. It can be seen graphically that on the vertical line, the real part is minimized at the saddle, while on the other, it is maximized. Hence we deform the contour to go along $(x,y(x))$, as shown in figure \ref{fig_contours_plane}. 

The integral along the deformed contour can be evaluated using the saddle approximation (including the determinant):
\begin{equation}
    \int dw e^{iqw + \pr{-\frac{q^2}{2\pi|\La|}+\frac{1}{2}}\ln\pr{w^2}} \approx e^{-i\frac{q^2}{2|\La|}} \pr{\frac{|q|}{\pi e|\La|}}^{-\frac{q^2}{\pi|\La|}} \frac{|q|}{\pi|\La|} \sqrt{\frac{2}{|\La|}}.
\end{equation}
Altogether the full integral is:
\begin{equation}
    \frac{2}{|\La|} \pr{\frac{|q|}{\pi e|\La|}}^{-\frac{q^2}{\pi|\La|}}e^{-i\frac{q^2}{2|\La|}}, 
\end{equation}
as the naive evaluation in (111) of \cite{Aharony:2023dod}.

\subsection{Torus}

We show here the two constant phase contours for the $\sg^1$ integral (\eqref{eq_sadddle_special_torus}, after setting $\tn = \sg^2 = 0$). One is along the imaginary axis, as can be seen immediately, and the real part is minimized at the saddle (see figure \ref{fig_sad}). The other is orthogonal to the imaginary axis, and the real part is maximized at the saddle. For $r<<1$, the deformed contour will look like the deformed contour for the plane, going upwards (see figure \ref{fig_im_re_plane}). However, for $r>>1$, the deformed contour goes downwards. 

The function $f$ for the torus, for $T_1=n_2=0$, was given in \eqref{eq_sadddle_special_torus}. The saddle $\sg^2 = \tn = 0$ holds for all $r$, and one remains with:
\begin{equation}
    f(w) = inw + \frac{nr}{2\pi}\pr{\ln{|\eta(\tw)|^6} -\ln\pr{\Th(w,\tw)^2}} - \frac{n}{4r} \frac{\pr{\tw + T_2}^2}{\tw},
\end{equation}
where $w$ is not $\sg^1 + \ttt\sg^2$, but rather the analytic continuation of $\sg^1$ to the complex cylinder. The part relevant for $w$ is:
\begin{equation}
    f(w) = iw - \frac{r}{2\pi}\ln\pr{\Th(w,\ttt)^2}.
\end{equation}
We removed $n$ since it is common for all factors. The saddles are the minima of $f$ along the imaginary axes. To present the different behavior for small and large $r$, we considered $T_2=10$, and $r = 0.3,10$. For these values, according to the analysis in section \ref{sec_sad}, one has $\ttt_{2s} \approx T_2,\frac{\pi T_2^2}{4r^2}$. The minima were shown in figure \ref{fig_sad}. The ``spikes" are the zeros of $\Th$, located at $\ttt_{2s}(r)$, with $\ttt_{2s} = 10,\ \pi/4$ respectively. For large $r$, one sees the minima due to the $\Th$-function only, which are approximately equally spaced, and the function presents an overall decrease. For small $r$, the first term $inw$ changes the location of the first two minima with respect to the others, being close to the right singularity rather than to the left singularity. 

\begin{figure}[!htb]
\captionsetup{singlelinecheck = false, justification=justified}
\includegraphics[scale=0.4]{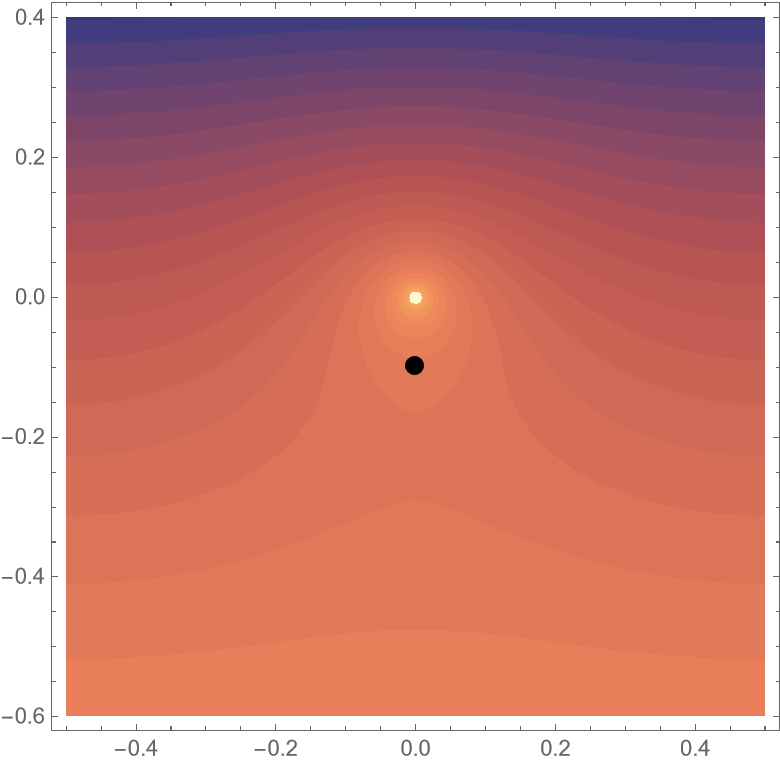}
\qquad
\includegraphics[scale=0.4]{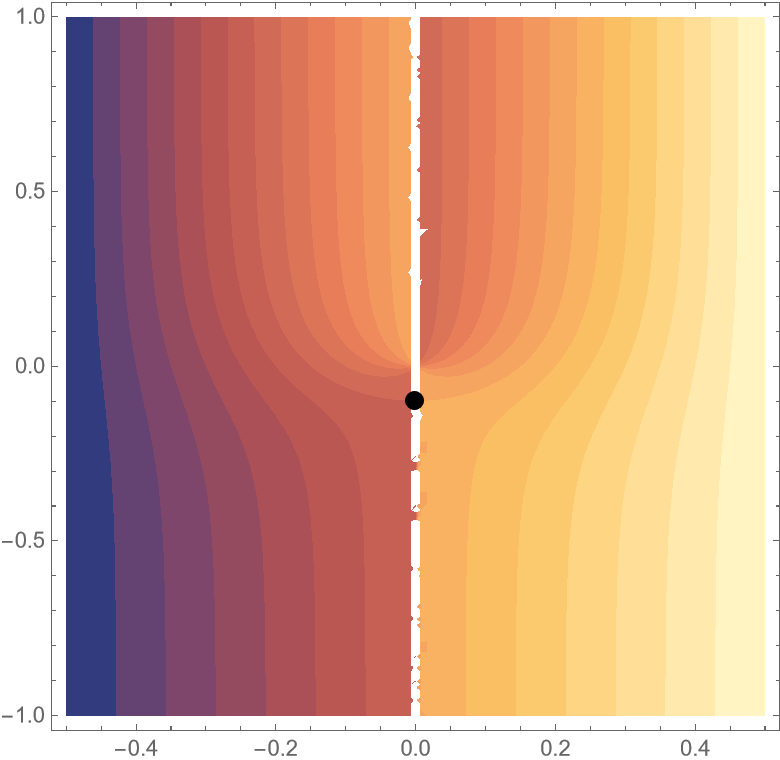}
\newline 
\includegraphics[scale=0.4]{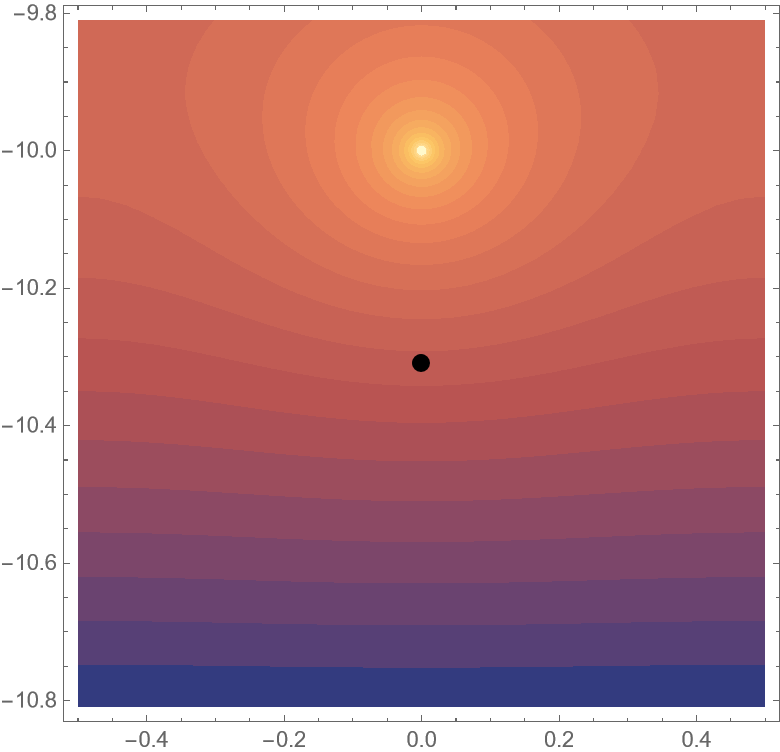}
\qquad
\includegraphics[scale=0.4]{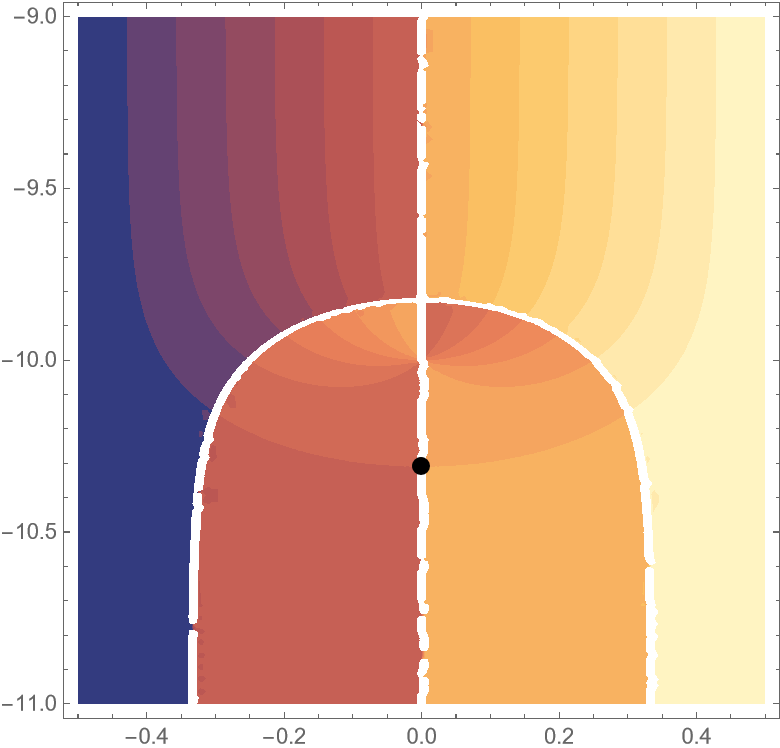}
\newline 
\includegraphics[scale=0.4]{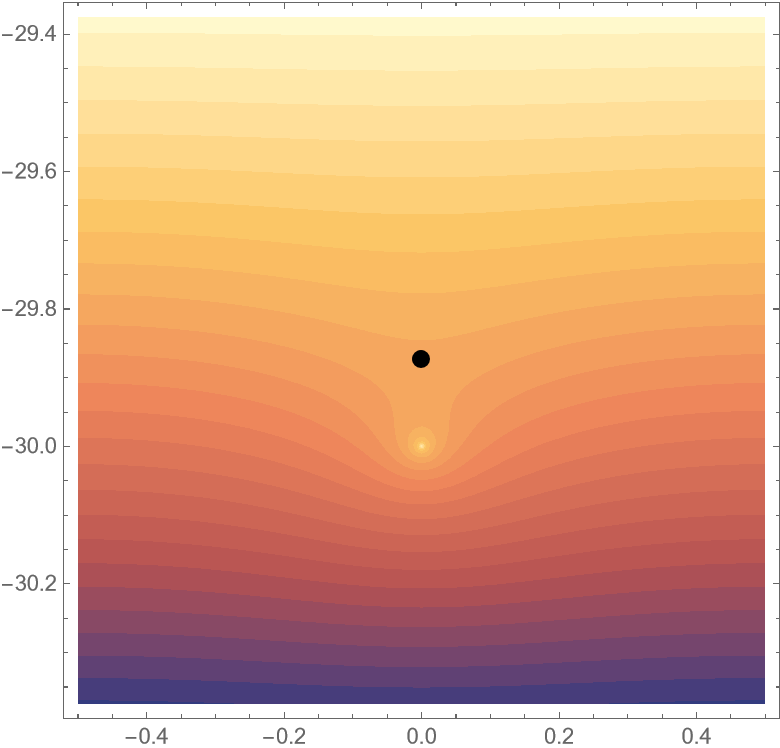}
\qquad
\includegraphics[scale=0.4]{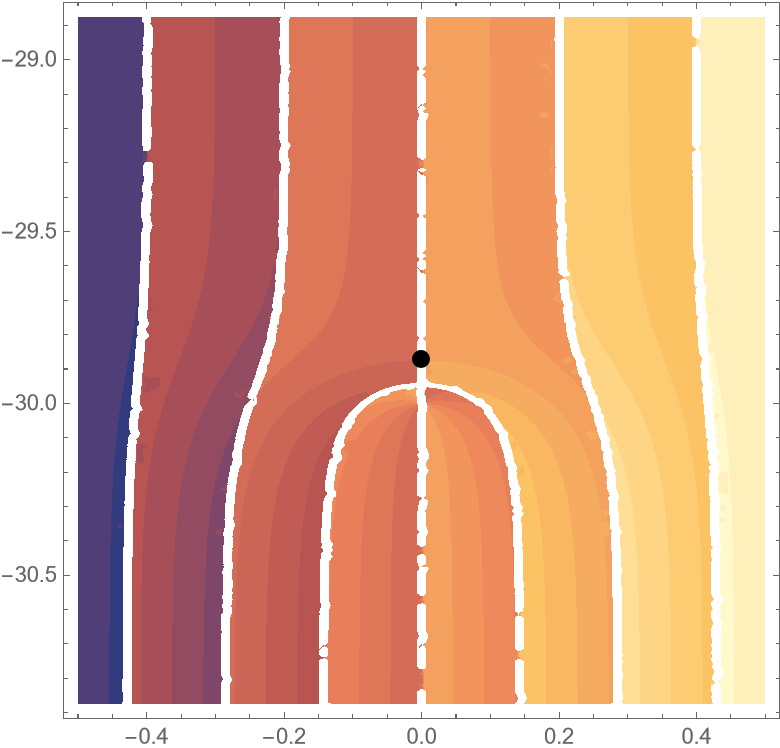}
\caption{Real (left) and imaginary (right) parts of $f$ around the first three saddles at small $r$. The saddles are located at the black point. One of the constant phase contour is the imaginary axis, the other can be seen on the figures. The bright points in the real part are the singularities due to the zeros of the $\Th$-function. The white lines in the imaginary parts are discontinuities due to the log. Note the different regions of the $y$-axis.}
\label{fig_saddle_torus_small_r}
\end{figure}

We show the real part and imaginary part of $f$ near its first three saddles for small $r$ in figure \ref{fig_saddle_torus_small_r}. The first will be relevant for the integral, but the third one, being dominated by the $\Th$-function, presents the large $r$ saddle. One sees that the constant phase contours of the first two are going up while the third one is going down. The contours for large $r$ were shown in figure \ref{fig_large_r_contours}, which resembles the third saddle of small $r$.


\end{document}